\documentclass[aip,pop,reprint,linenumber,superscriptaddress,amsmath,amsfonts]{revtex4-1}

\usepackage{amsfonts}

\usepackage[dvips]{color,graphicx}
\usepackage[colorlinks,linkcolor=blue,citecolor=blue,bookmarks,pdfstartview=FitH]{hyperref}
\usepackage{graphics}
\usepackage{ifthen}

\graphicspath{{figure/}}

\newcommand{\plotfigure}{true}

\begin{document}

\title{Theoretical and numerical studies of wave-packet propagation in tokamak plasmas}
\author{Z. X. Lu}
\email{luzhixin@pku.edu.cn}
\affiliation{State Key Laboratory of Nuclear Physics and Technology, School of Physics and Fusion Simulation Center, Peking University, Beijing 100871, China}
\affiliation{Associazione EURATOM-ENEA sulla Fusione, CP 65-00044 Frascati, Roma, Italy}
\author{F. Zonca}
\affiliation{Associazione EURATOM-ENEA sulla Fusione, CP 65-00044 Frascati, Roma, Italy} \affiliation{Institute for Fusion Theory and Simulation,
Zhejiang University, Hangzhou 310027, China}
\author{A. Cardinali}
\affiliation{Associazione EURATOM-ENEA sulla Fusione, CP 65-00044 Frascati, Roma, Italy}

%\date{}
%%%%%%%%%%%%%%%%%%%%%%%%%%%%%%%%%
\begin{abstract}
Theoretical and numerical studies of wave-packet propagation are presented to analyze the time varying 2D mode structures of electrostatic
fluctuations in tokamak plasmas, using general flux coordinates. Instead of solving the 2D wave equations directly, the solution of the initial value
problem is used to obtain the 2D mode structure, following the propagation of wave-packets generated by a source and reconstructing the time varying
field. As application, the 2D WKB method is applied to investigate the shaping effects (elongation and triangularity) of tokamak geometry on the
lower hybrid wave propagation and absorbtion. Meanwhile, the mode structure decomposition (MSD) method is used to handle the boundary conditions and
simplify the 2D problem to two nested 1D problems. The MSD method is related to that discussed earlier by Zonca and Chen [Phys. Fluids B 5, 3668
(1993)], and reduces to the well-known ``ballooning formalism'' [J. W. Connor, R. J. Hastie, and J. B. Taylor, Phys. Rev. Lett. 40, 396 (1978)], when
spatial scale separation applies. This method is used to investigate the time varying 2D electrostatic ITG mode structure with a mixed WKB-full-wave
technique. The time varying field pattern is reconstructed and the time asymptotic structure of the wave-packet propagation gives the 2D eigenmode
and the corresponding eigenvalue. As a general approach to investigate 2D mode structures in tokamak plasmas, our method also applies for
electromagnetic waves with general source/sink terms, either by an internal/external antenna or nonlinear wave interaction with zonal structures.
\end{abstract}

\maketitle

%%%%%%%%%%%%%%%%%%%%%%%%%%%%%%%%%

\section{Introduction and motivation}
\label{sec:intro} The two dimensional mode structure (in the radial and poloidal directions) in tokamak plasmas and its spatio-temporal evolution is
important since it is related to many physics aspects of fusion interest. For investigation of Alfv\'en
waves\cite{zonca92,zonca93,cheng95,vlad95,chen95,zonca96} and drift waves\cite{romanelli93,taylor93,connor93,taylor96}, the poloidal variation of the
plasma equilibrium gives rise to ``ballooning-like'' parallel mode structures, while the instability of the modes, in turn, may set the maximum
$\beta$ attainable in a tokamak\cite{connor78}. On the other hand, the radial fluctuation structure, e.g. the radial envelope of the mode, is related
to the radial correlation length of the underlying turbulence and might impact turbulent transport in various ways. Meanwhile, for radio-frequency
(RF) wave heating/current drive, the 2D mode structure, which bears the information of wave vector, wave intensity and location of the
reflection/resonant/mode conversion layers, is crucial for calculating the heating/current drive efficiency and deposition profile. For the case of
the lower hybrid wave propagation in tokamak plasmas, the toroidal effect, i.e. the breakdown of the poloidal symmetry of the system, causes a shift
of the parallel wave number and its evolution\cite{ignat:pop81toroidal,bonoli:pop82toroidal}, which is related to the current drive efficiency and
the wave energy deposition layer.

Besides the spatial structure, the temporal evolution of the fluctuations also incorporates important features. For example, the Ion Temperature
Gradient (ITG) mode, modulated by zonal flow, shows regular or chaotic behaviors, depending on the relative ordering of different characteristic
times, e.g. the inverse linear growth rate and the libration/rotation period of the wave-packet propagation\cite{zonca:pop04NL_interplay,zonca04}.
Zonal flows are also known to be connected with ITG turbulence spreading\cite{hahm:ppcf2004spreading, diamond:ppcf2005zf_review}, which may occur via
soliton formation\cite{guozh:prl09soliton}. These findings are helpful to understand basic features of transport process, e.g. the transport scaling
with the system size\cite{lin:prl2002scaling,diamond:ppcf2005zf_review, hahm:ppcf2004spreading}.

Although  mode structures can be readily obtained by many numerical solvers, either by eigenvalue approach, e.g. ERATO\cite{Gruber:cpc81erato} and
MARS\cite{bondeson:pfb92mars}, or solving the initial value problem as, e.g., in GTC\cite{lin:pop2000} and in the flux-tube simulations\cite{beer95},
analytical and semi-analytical techniques are developed to simplify the original problem, while maintaining the key physics. The ballooning formalism
makes use of the ``translational invariance'' of poloidal harmonics in the Fourier decomposition of the fluctuations, in order to reduce the 2D
eigenvalue problem to a local 1D parallel eigenvalue problem; thus, it gives the local eigenvalue and the parallel (to $\bf{B}$) mode
structure\cite{coppi77,lee77,glasser77,pegoraro78,connor78,connor79}. At the next order, the ballooning formalism takes into account the mode radial
envelope and the radial structure can be also obtained\cite{dewar81,dewar82,dewar83,zonca92,zonca93,zonca96,cheng95,vlad95,chen95,romanelli93,
taylor93, connor93,taylor96}. While these methods are based on the peculiar properties of MHD and drift wave fluctuations in strongly magnetized
plasmas, the WKB approach reduces the problem to the first order partial differential equation of the eikonal, so that the phase and the amplitude of
the wave can be obtained easily along the characteristic lines in weakly non-uniform systems (see
[\!\!\citenum{bernstein:pop1975,Weinberg:PhysRev1962}] for a review).

In this paper, we formulate the ``initial-value'' problem of wave-packet propagation  with a source term to get the mode structure in general tokamak
geometry using flux coordinates.  Rather than solving the partial differential equation for the wave directly, we can ``follow'' the evolution of
wave-packets generated by the source and calculate fluctuation patterns generated by them till the mode structure sets up. Here, the ``source'' can
be generic; for example, in the case of RF wave propagation, the source is provided in the form of the initial phase and amplitude determined by the
antenna located at the boundary\cite{brambilla82,bonoli:pop82toroidal}. In addition to this ``external'' antenna, the source term can represent an
``internal antenna'', which can be used to investigate mode structure, frequency and damping rate as, e.g., in the case of Alfv\'en
Eigenmodes\cite{deng:pop2010rsae,wang:ppcf10,zhang:pop11bae}. By adding the antenna induced perturbed scalar potential  $\delta\Phi_{ant}({\bf
r},t)=A_{ant}(r)cos(n_{ant}\phi-m_{ant}\theta)$ into the original equation, where $\phi$, $\theta$, $n$, $m$ denote the toroidal and poloidal angles
and the corresponding mode numbers, respectively, and the subscript ``ant'' stands for ``antenna'', the eigenmode is excited and the saturated wave
amplitude is given by\cite{Harris:handbook2002}
\begin{equation}\delta\Phi_{sat}\propto\frac{1}{\sqrt{(\omega_0^2-\omega^2_{ant})^2+4\gamma^2\omega_{ant}^2}}\;\;,\end{equation}
where $\omega_{ant}$ is the ``antenna'' frequency, $\omega^2_0=\omega_r^2+\gamma^2$, $\omega_r$ is the real frequency of the eigenmode and $\gamma$
is the damping rate. With all the information provided by the antenna excitation, we can reconstruct the mode structure in this region. In this
paper, we will discuss the general two dimensional propagation of wave-packets with an ``internal antenna'' inside the plasma or an ``external
antenna'' at the boundary. Generally, the source can also account for nonlinear wave interactions with zonal structures
\cite{zonca:nf2005transition,guozh:prl09soliton,chen:prl2004_zf,zonca:pop04NL_interplay} and, in the framework of general coordinates, it is
straightforward to extend this approach to three dimension when the toroidal symmetry breaks down.

The paper is organized as follows. In Section \ref{sec:coordmsd}, we introduce the general flux coordinate system and mode structure decomposition
(MSD) approach. There, we discuss the general use of MSD to represent the problem in the mapping space and to obtain the physics solution in real
space from that in the mapping space. The mixed WKB-full-wave approach for solving the 2D problem is also discussed along with its connection to the
ballooning formalism. In Section \ref{sec:WKB}, we introduce the WKB method for studying wave-packet propagation in general geometry, presenting
 its application to the propagation of cold lower hybrid waves. Meanwhile, the ITG eigenmode formation using the mixed
WKB-full-wave approach is analyzed and demonstrated. In Section \ref{sec:end}, we give our conclusions and final discussions. Two appendices are
devoted to more formal discussions of the mode structure decomposition approach and its connection with the well known ``ballooning formalism''.

\section{General flux coordinates and the mode structure decomposition approach}
\label{sec:coordmsd} In this paper, we use the general magnetic flux surface coordinates for our analyses of wave-packet propagation in plasmas of
fusion interest, characterized by complex geometries. Some of our discussion follows R. White's treatment of straight field line
coordinates\cite{white:springer2001}. We use $r,\hat\theta,\zeta$ to describe the radial-like, poloidal-like and toroidal-like coordinates, of which
the latter two have $2\pi$ periodicity. The radial-like variable $r$ is defined
by\begin{equation}\frac{r}{a}=\sqrt{\frac{\Psi-\Psi_0}{\Psi_b-\Psi_0}}\;\;,
\end{equation}
or \begin{equation}\frac{r}{a}=\sqrt{\frac{\Psi_T-\Psi_{T0}}{\Psi_{Tb}-\Psi_{Tb}}}\;\;,\end{equation} where $a$ is the normalization length, e.g. the
tokamak minor radius, $\Psi$ and $\Psi_T$ are poloidal and toroidal flux, respectively, while the subscripts ``$b$'' and ``$0$'' stand for the
boundary and on-axis value, respectively. In order to obtain the straight field line coordinates $(r, \hat\theta, \zeta)$, we shift $\zeta$ with
respect to the toroidal coordinate $\phi$ by $\nu(r,\hat\theta)=\phi-\zeta$\cite{white:springer2001} so that
\begin{equation}\label{q_straight_b}
\frac{{\bf B}\cdot\nabla\zeta}{{\bf B}\cdot\nabla\hat\theta}=q(r)\;\;,
\end{equation}
where  $\nu$ is a periodic function of $\hat\theta$. Here, the toroidal symmetry is preserved for $\zeta$ since
\begin{equation}
\left.\frac{\partial}{\partial\zeta}\right|_{r,\hat\theta}=\left.\frac{\partial}{\partial\phi}\right|_{r,\hat\theta}\;\;,\nonumber
\end{equation}
according to the chain rule. The choice of $\hat\theta$ is used to choose a convenient form for the Jacobian
\begin{equation}\label{jacobian_def}
J=(\nabla r\times\nabla\hat\theta\cdot\nabla\zeta)^{-1}=\frac{\partial{\bf  r}}{\partial r}\times\frac{\partial{\bf
r}}{\partial\hat\theta}\cdot\frac{\partial{\bf  r}}{\partial\zeta}\;\;.
\end{equation}
Examples are Hamada coordinates
\begin{equation}
J=J_H(r)
\end{equation}
and Boozer coordinates
\begin{equation}
J=J_B(r)/B^2\;\;,
\end{equation}
where the Jacobian or the product of the Jacobian with $B^2$ is a function of $\Psi$, respectively. Using equation (\ref{q_straight_b}), we can
obtain straight field line coordinates where the magnetic field can be written as
\begin{equation}\label{b_straight}
{\bf B}=\nabla\zeta\times\nabla\Psi-q(r)\nabla\hat\theta\times\nabla\Psi\;\;.
\end{equation}
With the definition\begin{equation} \xi=\zeta-q(r)\hat\theta\;\;,
\end{equation}
we can obtain the Clebsch representation for the magnetic field
\begin{equation} {\bf B}=\nabla\xi\times\nabla\Psi
\end{equation}
in the Clebsch coordinates $(r, \hat\theta, \xi)$\cite{beer95} or $(r, \hat\xi, \zeta)$\cite{Waltz:pop93stellarator}, with $\hat\xi=-\xi/q$,
depending on the choice of the parallel coordinate along the magnetic field line as $\hat\theta$ or $\zeta$ and the rescaling of $\Psi$ and $\xi$,
which are stream functions of the magnetic field.

Straight field line coordinates have good features, such as constant safety factor on magnetic flux surfaces and the advantage of easily describing
the parallel (to $\bf B$) mode structure. In addition, different coordinate systems represent different features. For example, $(r, \hat\xi, \zeta)$
can handle the geometry with $q=\infty$ (X-point configuration)\cite{Waltz:pop93stellarator}, where the $(r, \hat\theta, \xi)$ coordinates fail since
$\xi=\zeta-q\hat\theta\sim -q\hat\theta$. Usually, the appropriate coordinate system is chosen according to the needs of the calculation. For
example, the gyro-kinetic turbulence code GTC makes use of straight field line coordinates to suitably describe the field-aligned structure of
micro-instability\cite{lin:pop2000} and the particles are pushed with guiding-center Hamiltonian formalism with conserved phase
volume\cite{white:pop84}. The grid points in the Clebsch coordinates $(r,\hat\xi,\zeta)$ are aligned to follow the equilibrium magnetic field lines and, thus, the number
of grid points aligned along the parallel coordinate can be greatly reduced since the modes have elongated structures along the field lines
($k_{||}\ll k_\perp$). As for the particle pushing, a simple Runge-Kutta scheme can be used safely and accurately to integrate the Hamiltonian system
 since particles move mainly along a straight line in the straight field line coordinate system $(r, \hat\theta, \zeta)$. In flux-tube
simulations\cite{beer95}, Clebsch coordinates $(r, \hat\theta, \xi)$ are used and the global domain is replaced with flux tubes along the magnetic
field line, $\Psi\in(\Psi_0-\Delta\Psi,\Psi_0+\Delta\Psi)$, $\xi\in(\xi_0-\Delta\xi,\xi_0+\Delta\xi)$,
$\hat\theta\in(\hat\theta_0-\Delta\hat\theta,\hat\theta_0+\Delta\hat\theta)$. The main concept to deal with the boundary condition is the
statistically motivated periodicity, which assumes that the statistical properties of fluctuations at the boundaries along $\Psi$, $\xi$ and
$\hat\theta$ are identical when the simulation domain is larger than the correlation length along the three coordinates. When the parallel
correlation length increases and becomes larger than $2\pi$, the parallel length of the box needs to be extended beyond $2\pi$ and non-physical
parallel wavelengths are permitted, which requires a dedicated discussion about the global consistency\cite{scott98}. Many works\cite{lin:pop2000,
beer95, white:springer2001, Waltz:pop93stellarator, white:pop84, scott98}, including recent ones \cite{ottaviani:PLA2011coordinates}, deal with the
formulation of field-aligned flux coordinates and their implementation in numerical codes; so, we do not discuss this issue further.

In our approach, in order to deal with boundary conditions, we make use of the mode structure decomposition (MSD) approach\cite{zonca04}, which can
be reduced to the well-known ``ballooning formalism''\cite{coppi77,lee77,glasser77,pegoraro78,connor78,connor79,hazeltine81,dewar81,dewar82} when
spatial scale separation applies between fluctuation structures and equilibrium non-uniformities. Writing the fluctuation in the form
\begin{equation}\label{eq:f_decompose}
f=e^{in\zeta}f_n(r,\hat\theta)=e^{in\zeta}\sum_m e^{-im\hat\theta}f_{n,m}(r)\;\;,
\end{equation}
 by continuing $f_{n,m}(r)$ to $\varphi_n(r,x)$, i.e. $\varphi_n(r,x)\in\{\varphi_n(r,x)|\varphi_n(r,m)=f_{n,m}(r),x\in\mathbb{R},m\in \mathbb{Z}\}$, and defining the Fourier transform
\begin{equation}\label{eq:hatf_construction}
\hat f_n (r,\eta) = (2\pi)^{-1} \int e^{-i\eta x} \varphi_n(r,x) d x \;\; ,
\end{equation} we have
\begin{eqnarray}\label{eq:periodization0}
f_n(r,\theta) & = & 2\pi \sum_\ell \hat f_n (r,\theta-2\pi\ell)\nonumber \\
& = & \sum_m e^{-im\theta} \int e^{im\eta} \hat f_n (r,\eta) d\eta \;\; ,
\end{eqnarray} where the Poisson summation formula has been used in the equivalent form of equation (\ref{eq:psf}) in Appendix \ref{app:msd}.
Then, $f_n(r,\hat\theta)$ can be obtained by solving the linear Partial Differential Equation (PDE)
\begin{equation} \label{eq:equation_mapping}\mathcal {L}(r,\eta;\partial_r,\partial_{\eta})\hat
f_n(r,\eta)=0\;\;,
\end{equation} with suitable boundary conditions in the mapping space and performing summation via equation (\ref{eq:periodization0}), rather than solving the original equation
\begin{equation}
\label{eq:equation_real}\mathcal{L}(r,\hat\theta;\partial_r,\partial_{\hat\theta})f_n(r,\hat\theta)=0\;\;,
\end{equation}
with $2\pi$ periodic boundary condition in $\hat\theta$ direction. In equations (\ref{eq:f_decompose}) to (\ref{eq:equation_real}), the toroidal
symmetry has been used for the reduction of the PDE corresponding to the considered wave equation from 3D to 2D. The partial differential operators
above are carried out holding the other coordinate constant as well as the third one, e.g. $\xi$ in the following Clebsch coordinates. The
representation of $\mathcal{L}(r,\eta;\partial_r,\partial_{\eta})$ can be obtained by mapping the differential operators in $\mathcal{L}$ from real
space to the mapping space. Since these operators remain the same in real and mapping space (equation (\ref{eq:mapping0}) in Appendix \ref{app:msd})
except the substitution of $\hat\theta$ with $\eta$, $\mathcal{L}$ remains formally the same in equation (\ref{eq:equation_mapping}) and
(\ref{eq:equation_real}). Although the solution in real space can be obtained from that in the mapping space, in other words, equation
(\ref{eq:equation_mapping}) implies equation (\ref{eq:equation_real}), the opposite is not true. For details, please see Appendix \ref{app:msd},
where equation (\ref{eq:diff1}) demonstrates this statement, which is connected with the non-uniqueness of the construction of $\hat
f_n(r,\hat\theta)$ from $f_n(r,\hat\theta)$ in equation (\ref{eq:hatf_construction}), for the sampling operator $\varphi_n(r,x) \mapsto f_{n,m}(r)$
admits different choices of $\varphi_n(r,x)$. Some previous works\cite{dewar83,hazeltine81,hazeltine90} have discussed the uniqueness of one specific
choice of the inverse ``ballooning'' transformation. However, generally speaking, the representation of equation (\ref{eq:hatf_construction}) is not
unique and different constructions of $\varphi_n(r,x)$ are possible, as discussed in Reference [\!\!\citenum{zonca04}] and Appendix \ref{app:msd}.

In practical applications, it is often useful to adopt Clebsch coordinates $(r,\hat\theta,\xi)$, where
\begin{equation}
f(r,\hat\theta,\xi)= \sum_n e^{i n \xi} F_n(r,\hat\theta) \;\; .
\end{equation}
From a  direct comparison with equation (\ref{eq:f_decompose}), we obtain
\begin{equation} \hat f_n (r,\eta) = e^{-inq\eta} \hat F_n(r,\eta) \;\; , \label{eq:hatbigf_sec2}
\end{equation}
from which, noting equation (\ref{eq:periodization0}), we have
\begin{eqnarray}
F_n(r,\theta) & = & 2\pi \sum_\ell e^{2\pi i \ell nq} \hat F_n (r,\theta-2\pi\ell)\nonumber \\
& = & \sum_m e^{i(nq-m)\theta} \int e^{i(m-nq)\eta} \hat F_n (r,\eta) d\eta \;\; . \label{eq:periodization2_sec2}
\end{eqnarray}
 The original differential equation, equation (\ref{eq:equation_real}), after transformation to Clebsch coordinates, can be written as
\begin{equation}
\label{eq:equation_realG}\mathcal{G}(r,\hat\theta;\partial_r,\partial_{\hat\theta})F_n(r,\hat\theta)=0\;\;,
\end{equation} in real space, or
\begin{equation}
\label{eq:equation_mappingG}\mathcal{G}(r,\eta;\partial_r,\partial_{\eta})\hat F_n(r,\eta)=0\;\;,
\end{equation}
in the mapping space, where the linear differential operators $\mathcal{G}$ are formally the same in the two equations. The solution
$F_n(r,\hat\theta)$ can be obtained by solving equation (\ref{eq:equation_mappingG}) with suitable boundary conditions and the construction by
equation (\ref{eq:periodization2_sec2}).

The mixed WKB-full-wave approach\cite{zonca04, cardinali03}, as generalization of the standard ballooning-mode formalism\cite{connor78}, can be used
to solve equation (\ref{eq:equation_mappingG}). When its applicability condition is satisfied, i.e. the wave propagates much faster in the parallel
than in the radial (magnetic flux) direction, the parallel mode structure is formed before the wave propagates significantly away from the considered
magnetic field line; thus, it is meaningful to separately calculate the parallel mode structure and the radial envelope. The mixed WKB-full-wave
approach, proposed here, is essentially the same as that used in [\!\!\citenum{dewar81,dewar82,dewar83,dewar:varenna87,dewar97}] for investigating
high mode number MHD modes and resulting remarkably good even down to low mode numbers. The same method was used later for analyzing eigenmode
stability of drift waves\cite{romanelli93,taylor93,connor93,taylor96} as well as Alfv\'enic
modes\cite{zonca92,zonca93,cheng95,vlad95,chen95,zonca96}. The main novelty of our present treatment stands in the solution of the wave equation as
initial value problem in the presence of a source term, which makes extensions to nonlinear problems readily available\cite{zonca04} and allows us to
present the wave-packet propagation in magnetized plasmas with general equilibrium geometries as one single coherent framework, with different
possible applications to propagation of radio-frequency (RF) waves in toroidal geometries (section \ref{sec:lhcold}), mode structure and stability of
drift waves (section \ref{sec:ITG}) as well as Alfv\'enic fluctuations and MHD modes. Since $\eta$ denotes the coordinate along the equilibrium
magnetic field, as an ``extended poloidal angle''\cite{connor78}, equation (\ref{eq:equation_mappingG}) gives the parallel wave equation by
substituting $\partial_r$ with $ik_r(r)$
\begin{equation}
\label{eq:equation_mappingG_hybrid}\mathcal{G}(r,\eta;ik_r,\partial_{\eta})\hat A(r)\hat F_{n0}(r,\eta)=0\;\;,
\end{equation}
where $\hat A(r)=e^{i\int\ k_rdr}$ is written in the eikonal form. Then, given the solution of the parallel wave equation, $k_r$ is obtained from
\begin{eqnarray}
D(\omega,r,k_r)\hat A(r)=0\;\;,
\end{eqnarray}
where $\omega$ is the mode frequency and the local dispersion relation $D(\omega,r,k_r)$ is obtained from the solution of equation
(\ref{eq:equation_mappingG_hybrid}) with proper boundary conditions\cite{dewar:varenna87,dewar97}. The radial envelope $\hat A(r)$ can be obtained
using the WKB method when $|\partial_rk_r/k^2_r|\ll1$ (see section \ref{sec:WKB}). The applicability condition of the mixed WKB-full-wave method is
more stringent than the obvious condition on the ratio of radial to parallel wave-packet group velocities being small;
$|v_{gr}/v_{g||}|\approx|k_{||}/k_{\perp}|\ll1$. The condition that the parallel mode structure is formed before any significant radial propagation
takes place is given by $|v_{gr}|\ll |v_{g||}L_T/L_{||}|$, with $L_T\sim r$ the characteristic arc-length of the wave-packet trajectory in a toroidal
plasma cross-section and $L_{||}\sim qR_0$ the connection-length of a tokamak of major radius $R_0$. Thus, the actual applicability condition of the
mixed WKB-full-wave method is $|v_{gr}/v_{g||}|\approx|k_{||}/k_\perp|\ll r/(qR_0)<1$. The mixed WKB-full-wave method can be used, e.g. to calculate
the 2D structure of TAE\cite{zonca93} and ITG modes\cite{romanelli93}. It has been also proposed to investigate the lower hybrid wave propagation
with $k_\parallel\ll k_\perp$, or more generally, with a small ratio of radial to parallel group velocities\cite{cardinali03}.  As an example of the
mixed WKB-full-wave approach, used for solving an initial value wave equation in the presence of a source term, we discuss the calculation of the 2D
electrostatic ITG mode structure in section \ref{sec:ITG}. Meanwhile, for illustrating the calculation of the time-evolving 2D wave structure of RF
waves in general toroidal geometry by means of 2D WKB method, we analyze the case of the propagation of a cold Lower Hybrid wave-packet in section
\ref{sec:lhcold}. The application of the mixed WKB-full-wave approach to the LH propagation requires a dedicated discussion and will be reported
elsewhere.

Before discussing specific applications of the mode structure decomposition (MSD) method, we give a brief discussion of its connection with the
ballooning formalism. A more formal and detailed analysis is given in Appendix \ref{app:ballooning}. As anticipated above, the mode structure
decomposition method is general and can be reduced to the ballooning formalism when spatial scale separation applies between fluctuation structures
and equilibrium non-uniformities. In the mapping space of Clebsch coordinates $(r,\hat\theta,\xi)$, when the equilibrium profile variation is much
less than that of the radial wavelength, then $|\partial_r\hat F(r,\eta)|\ll|nq'\hat F_n(r,\eta)|$. This condition is equivalent to the quasi
translational invariance of poloidal harmonics in ballooning formalism, where $f_{n,m}(r)=A(r)\bar f_n(nq+m)$, with $A(r)$ being the slowly varying
amplitude. Thus using $\hat F_n(r,\eta)=A(r)\bar F_n(r,\eta)$, we have the ballooning representation
\begin{eqnarray}
\label{eq:ballooning_cleb} f_n(r,\hat\theta) & = & A(r)\sum_m e^{-im\hat\theta} \int e^{i(m-nq)\eta} \bar F_n (r,\eta) d\eta \;\; ,
\end{eqnarray} where
$A(r)$ and $\bar F_n (r,\eta)$ vary on the envelope scale $L_A$ and the profile characteristic length $L_p$, respectively, with $L_A\ll L_p$. Using
the scale separation argument, we can also get the inverse of ballooning transformation proposed by Hazeltine
[\!\!\!\citenum{hazeltine81,hazeltine90}], which makes the mapping between $f_n(r,\eta)$ and $\hat f_n(r,\hat\theta)$ to be one-to-one. In Appendix
\ref{app:ballooning}, we demonstrate that equation (\ref{eq:equation_mapping}) is the necessary and sufficient condition for equation
(\ref{eq:equation_real}) to be satisfied, with proper scale separation and using Hazeltine's prescription for construction of the image function in
the mapping space.  We also clarify the more general situation in the framework of the MSD approach.

An advantage of moving to the mapping space to solve the equation for $\hat f_n(r,\eta)$ or $\hat F_n(r,\eta)$ is that we can easily choose proper
boundary condition in $\eta$ direction, rather than the $2\pi$ periodic constraint in $\hat\theta$ direction. Typical boundary conditions are
outgoing wave, where no energy is generated at large $\eta$. When the ``flux tube'' is chosen long enough, the vanishing boundary condition can be
chosen as well. Then, with the solution in the mapping space, the physical solution satisfying periodicity constraints in $\hat\theta$ direction can
be uniquely constructed with equations (\ref{eq:periodization0}) or (\ref{eq:periodization2_sec2}). In addition, since the MSD approach only relies
on the Poisson summation formula, it is not limited by the condition $q'\ne0$ required by the ballooning formalism, and its application to cases with
vanishing magnetic shear is readily obtained\cite{zonca02,connor04}.

\section{Complex WKB formulation of the wave-packet propagation problem in general toroidal geometry}
\label{sec:WKB} The ``WKB'' is a method for calculating approximate solutions of linear partial differential equations with varying coefficients. In
plasma wave propagation, it is applied to investigate the solution of the Maxwell-Vlasov system . Based on the eikonal form of the solution of the
wave equation
\begin{equation}\label{eq:eikonal} \Phi({\bf  r})=A\exp\{i\tilde{S}_0\}\;\;,
\end{equation}
it relies on the geometric optics approximation, i.e. the ratio between wavelength and the characteristic length of the variation of the reflective
index must be small, \begin{equation}\label{eq:geometric_approx}\epsilon=\left|\frac{\nabla\cdot{\bf  k}}{k^2}\right|\ll1\;\;.\end{equation} By using
the ansatz equation (\ref{eq:eikonal}) and (\ref{eq:geometric_approx}), the ray trajectory equation system and the amplitude equation are obtained
from the asymptotic expansion in $\epsilon$ of the governing wave equation. This set of first order differential equations describe, at the lowest
order, the wave propagation and, at the higher order, the focusing/defocusing effects of the propagating rays.  While traditional geometrical optics
deals with the diffractiveless wave fields, beam tracing methods were developed to investigate the diffraction of lower hybrid wave
propagation\cite{pereverzev:nu92}. The beam tracing method, even though it preserves the Hamiltonian particle description for the wave in the group
velocity ${\bf  v}_g$ direction as the usual WKB method, retains the full wave description in the direction perpendicular to ${\bf  v}_g$; the
additional ordering between the wavelength and wave beam size is then used to expand the original wave equation as asymptotic series. The beam
tracing is a complex WKB method with extra ordering suited to describe the finite beam width, which means that the physical phenomenon of diffraction
is included in the treatment of the wave equation. The general complex geometric method for investigating the Gaussian beams in inhomogeneous media
is reviewed by Yu. Kravtsov\cite{kravtsov:springerlink2007}. Here, we'll apply it to describe plasma waves-packet propagation and obtain the
reduction to the traditional WKB method when the correction from diffraction is negligible. As anticipated in Section \ref{sec:coordmsd}, we will
also discuss the conditions under which wave-packet propagation in 2D system can be studied with a mixed WKB-full-wave approach.

Using WKB method, the solution of the wave equation is constructed along the characteristic lines, parallel to the wave group velocity, along which
the energy is transported. We can apply it in real space directly or in the mapping space, after decomposing the wave-packet with the mode structure
decomposition (MSD) method (see Section \ref{sec:coordmsd}). According to the results of section \ref{sec:coordmsd}, the equations in both spaces are
equivalent and formally the same. However, the different boundary conditions lead to different solutions connected with periodic operators (see
Appendix \ref{app:msd} and \ref{app:ballooning}).

Here, we discuss how to reconstruct the time-varying wave field by the WKB method. In the electrostatic limit, the wave equation reduces to
\begin{equation}\label{eq:L=0}
\mathcal{L}({\bf  r},\nabla)\Phi({\bf  r})=0\;\;,
\end{equation}
where $\Phi$ is the perturbed scalar potential and the operator $\mathcal{L}({\bf  r},\nabla)$ is determined by the plasma parameters and the
structure of the dielectric tensor. Assuming $\nabla\Phi({\bf  r})=i\left\{\nabla\tilde S({\bf  r})\right\}\Phi({\bf  r})=i{\bf  k}\Phi({\bf  r})$
and solving equation (\ref{eq:L=0}) with an asymptotic expansion in $\epsilon$, we have
\begin{equation}\label{eq:localDR}
D_0({\bf  r},{\bf  k}_0)\Phi({\bf  r})\equiv \mathcal{L}({\bf  r},i{\bf  k}_0)\Phi({\bf  r})=0
\end{equation}
at the lowest order $O(\epsilon^0)$, where we have denoted with $\bf k_0$ the first term of the asymptotic series ${\bf  k}={\bf  k}_0+{\bf k}_1+{\bf
k}_2\ldots$. Equation (\ref{eq:localDR}) is readily solved by the ray tracing equation system (method of characteristics):
\begin{eqnarray}
\label{eq:drdtau0}&&\frac{d{\bf r}}{d\tau}=-\frac{\partial D_0}{\partial{\bf k}_0}\;\;,\\ \nonumber &&\frac{d{\bf k}_0}{d\tau}=\frac{\partial
D_0}{\partial{\bf r}}\;\;,
\end{eqnarray}where $\tau$ is a time-like coordinate parameterizing the wave-packet motion along the ray trajectory. Then, the lowest order eikonal in
equation (\ref{eq:eikonal}) can be obtained by integration along the trajectory
\begin{equation} \tilde S_0=\int{\bf k}_0\cdot \frac{d{\bf r}}{d\tau}d\tau\;\;.
\end{equation}
Meanwhile, ${\bf k}_1$, ${\bf k}_2\ldots$ can be obtained by expanding the wave equation at the higher order. The ${\bf k}_1$ correction corresponds
to the leading order amplitude term in the eikonal representation of equation (\ref{eq:eikonal}) and is given by
\begin{equation}
A({\bf r})=\exp\{i\int{\bf k}_1\cdot\frac{d{\bf r}}{d\tau}d\tau\}\;\;.
\end{equation}Knowing $A$ and $\tilde S_0$, it is possible to reconstruct the leading order wave field accurate to $O(\epsilon^{1})$.

In equation (\ref{eq:drdtau0}), the complex ray tracing method takes $D_0$ as a complex function and thus, the imaginary part of $\partial
D_0/\partial{\bf k}_0$ leads to imaginary displacement of the ray in the complex space. In the framework of complex ray tracing, rays have a real
physical meaning only when they intersect the real space. Thus, rays should be started with given boundary condition in complex space, analytically
continued from the real boundary condition, and propagate in the complex space till they reach the real space\cite{kravtsov:springerlink2007}. Unlike
beam tracing, where an additional small parameter, the ratio between wavelength and beam width, is introduced to expand the wave equation and to
force the beam to move in real space, the complex WKB method deals with the ray equation in complex space without additional scale
separation\cite{mazzucato:pop1989,kravtsov:springer1990}. Using complex WKB method, diffraction can be described well and a heuristic result in
homogeneous media shows that the critical propagation length for diffraction is
\begin{equation} L_c=\frac{W^2}{\lambda}\;\;,
\end{equation}
where $W$ is the Gaussian beam width and $\lambda$ is the wavelength. For propagation length less than $L_c$ the diffraction effect is negligible and
traditional WKB method works reasonably well.

Coming to the case of tokamak plasmas, considering the equilibrium toroidal symmetry, linear mode structures can be Fourier decomposed in the
toroidal direction and, for a given toroidal mode number $n$, the problem is reduced to two dimensional and equation (\ref{eq:L=0}) reduces to
\begin{equation} \mathcal{L}(r,\hat\theta,\partial_r,\partial_{\hat\theta})\Phi_n(r,\hat\theta)=0\;\;, \end{equation}
where we assumed $\Phi({\bf r})=e^{in\zeta}\Phi_n(r,\hat\theta)$. To study this equation, we can generally use a 2D WKB method, as we will do in
section \ref{sec:lhcold} for the case of a cold lower hybrid wave-packet; or we can adopt the mixed WKB-full-wave approach, introduced in section
\ref{sec:coordmsd}, as in the application to the electrostatic ITG propagation discussed in section \ref{sec:ITG}.  In the following section
\ref{sec:lhcold}, we will employ the traditional WKB method to discuss the case of the lower hybrid wave propagation in the electrostatic limit,
reconstructing the time dependent wave-field pattern in 2D geometry. Meanwhile, to demonstrate the mixed WKB-full-wave method, we discuss its
application in the case of the electrostatic ITG propagation and eigenmode formation in torus in section \ref{sec:ITG}. The application of the this
method to the lower hybrid wave propagation will be reported elsewhere, since it requires a dedicated work to discuss its applicability and to
compare its results with the findings from a more conventional 2D WKB method.

%%%%%% section 3B LH
\subsection{Propagation of a cold lower hybrid wave-packet}
\label{sec:lhcold}

As the first application of the method illustrated above, we consider the propagation of the cold lower hybrid wave in the frequency range
$\omega_{ci}\ll\omega\ll\omega_{ce}$. As a quasi-electrostatic wave, lower hybrid wave can be described by the Poisson equation
\begin{equation}\label{eq:poisson}
\mathcal{L}({\bf r},\nabla)\Phi({\bf r})=\nabla \cdot \big(\underline{\underline{\varepsilon}}({\bf r})\cdot \nabla\Phi({\bf r})\big)=0\;\;,
\end{equation}
where $\underline{\underline{\varepsilon}}({\bf r})=S \textit{\underline{\underline{I}}} + (P-S) {\bf b}{\bf b} - i D {\bf b}\cdot\tilde\varepsilon $
is the cold dielectric tensor, $\tilde\varepsilon$ is the Levi-Civita tensor, and $S$,$D$,$P$ are elements of the cold dielectric tensor in Stix
notation. Thus we have
\begin{equation}
\nabla \cdot \left(S\nabla\Phi({\bf r})\right)+\nabla \cdot \left((P-S)\nabla_\parallel\Phi({\bf r})\right)=0
\end{equation}
and correspondingly,
\begin{eqnarray}
\label{eq:D0=0}D_0({\bf r},{\bf k}_0) & \equiv & Sk_0^2+(P-S)k^2_{\parallel 0}  = 0\;\;,\\
\label{eq:D1=0}D_1({\bf r},{\bf k}_1) & \equiv & -i\nabla\cdot(P{\bf k}_{\parallel 0}+S{\bf k}_{\perp 0})\\ \nonumber
 & &+2(P{\bf k}_{\parallel 0}\cdot{\bf k}_{\parallel 1}+S{\bf k}_{\perp 0}\cdot{\bf k}_{\perp 1})=0
\end{eqnarray}
in the zeroth and first order WKB expansion. In the straight field line coordinates $(r,\hat\theta,\zeta)$ mentioned in section \ref{sec:coordmsd},
equation (\ref{eq:D0=0}) can be written as
\begin{equation}\label{eq:D0}
D_0=Sg^{\alpha\beta}k_{\alpha 0} k_{\beta0}+\frac{P-S}{J^2B^2}(nq+m)^2=0\;\;,
\end{equation} where $\alpha,\beta\in \{r,\hat\theta,\zeta\}$ and $k_{\alpha 0}$, $k_{\beta 0}$ are their conjugate momenta defined by $k_{\alpha
0}=\partial S_{0}/\partial \alpha$, $k_{\beta 0}=\partial S_{0}/\partial \beta$. The metric elements are defined as $g^{\alpha\beta}={\bf
e}^\alpha\cdot{\bf e}^\beta$ with the contra-variant basis ${\bf e}^\alpha=\nabla\alpha$.  The repeated superscript and subscript mean summation.
Then the ray tracing equations are
\begin{eqnarray}
\label{eq:drdtau}\frac{dr}{d\tau} & = & -2Sg^{r\beta}k_{\beta 0}\;\;, \\
\label{eq:dthetadtau}\frac{d\hat\theta}{d\tau} & = & -2\left[Sg^{\hat\theta\beta}k_{\beta 0}+(P-S)\frac{nq+m}{J^2B^2}\right]\;\;,\\
\label{eq:dzetadtau}\frac{d\zeta}{d\tau} & = & -2\left[Sg^{\zeta\beta}k_{\beta 0}+(P-S)\frac{nq+m}{J^2B^2}q\right]\;\;,\\
\label{dkdtau}\frac{dk_{\alpha 0}}{d\tau} & = & \left[\left(\frac{\partial}{\partial\alpha}S\right)g^{\beta\gamma}
+\left(\frac{\partial}{\partial\alpha}g^{\beta\gamma}\right)S\right]k_{\beta0}k_{\gamma0} \\ \nonumber
&&+\left\{\frac{1}{P-S}\frac{\partial (P-S)}{\partial\alpha}+2\left[\frac{n}{nq+m}\frac{\partial}{\partial\alpha}q\right.\right. \\ \nonumber
&&\left.\left.-\frac{1}{JB}\frac{\partial}{\partial\alpha}(JB)\right]\right\}(P-S)\frac{(nq+m)^2}{J^2B^2}\;\;.
\end{eqnarray}

The initial condition is given at $\tau=0$ on the initial surface $Q$, parametrically defined as ${\bf r}={\bf r}_I(\iota,\varsigma)$ where $\iota$,
$\varsigma$ are curvilinear coordinates on $Q$ and subscript $I$ means initial value. In tokamak, we can choose $\iota=\hat\theta$, $\varsigma=\zeta$
at the starting surface ${\bf r}={\bf r}(r=r_I,\hat\theta,\zeta)$. Then the initial condition of eikonal $S_{0I}$ and amplitude $A$ is
\begin{eqnarray}
S_0|_Q=S_{0I}(\hat\theta,\zeta)\;\;,\\
A_0|_Q=A_{0I}(\hat\theta,\zeta)\;\;,
\end{eqnarray}
and $(k_r,m,n)|_Q$ are obtained from $S_{0I}$ by their definition and local dispersion relation.

At the higher order, equation (\ref{eq:D1=0}) for the amplitude can be written as
\begin{equation}
\frac{d A}{d \tau}+\frac{A}{2}\nabla\cdot\frac{d {\bf r}}{d \tau}=0\;\;,
\end{equation}
where \[\frac{d A}{d \tau}=\frac{d{\bf r}}{d\tau}\cdot\nabla A \;\;,\] \[\frac{d{\bf r}}{d\tau}=-\frac{\partial D_0}{\partial{\bf k}_0}\;\;.\] To
derive $\nabla\cdot\frac{d {\bf r}}{d \tau}$, we move to the ray coordinates $(\iota,\varsigma,\tau)$, where $\iota$,$\varsigma$ label the rays
starting from the initial surface $Q$ and $\tau$ indicates the distance along a fixed ray. In this coordinate, the amplitude is
\begin{equation}
 A(\tau)=A(\tau_0)\sqrt{\frac{J_R(\tau_0)}{J_R(\tau)}}\;\;,
\end{equation}
where the Jacobian of the ray coordinates is defined as $J_R=\frac{\partial{\bf r}}{\partial\tau}\times\frac{\partial{\bf
r}}{\partial\eta}\cdot\frac{\partial{\bf r}}{\partial\xi}$\cite{kravtsov:springer1990}. Applying this result to the tokamak geometry and making use
of equation (\ref{jacobian_def}), we have
\begin{equation}
 J_R=J\left(\frac{\partial r}{\partial\tau}\frac{\partial\hat\theta}{\partial\iota}-\frac{\partial\hat\theta}{\partial\tau}\frac{\partial
r}{\partial\iota}\right)\;\;,
\end{equation} where we make use of the toroidal symmetry to simplify $J_R$ since $\frac{\partial\alpha}{\partial\varsigma}=\delta_{\alpha\varsigma}$.
Here $\delta_{\alpha\varsigma}$ is the Kronecker delta. This is also the formula implemented in the numerical calculation.

Although the equation set is quite general, we can still get some qualitative information on the properties of its solutions, such as on the location
of the reflection points where $dr/d\tau=0$. From equation (\ref{eq:drdtau}), we have \[\frac{dr}{d\tau}=-2S{\bf e}^r\cdot{\bf e}^\perp
k_{\perp0}\;\;,\] where ${\bf e}^\perp$ is the perpendicular unit vector. Thus two types of reflection points exist. One appears near the cut off
layer with $k_{\perp0}\rightarrow0$. The other appears usually in the inner region where ${\bf e}^r\cdot{\bf e}^\perp=0$. The lower hybrid wave, once
it has been efficiently coupled with the plasma from the external launching system, is trapped between the two reflection points until absorbed by
electron Landau damping (ELD). This phenomenon is referred to as ``multi-reflection''\cite{brambilla82}. We can also assume simplified tokamak
geometry, e.g. concentric circular magnetic surfaces, to reduce the equations and obtain asymptotic solutions of the wave propagation
semi-analytically or numerically. This is investigated and discussed in Ref. [\!\!\citenum{cardinali07}].

In order to calculate the propagation and absorption of the lower hybrid wave in a general self-consistent plasma equilibrium, the magnetic field
$B$, the Jacobian $J$ and the metric tensor $g^{ij}$ are calculated for a given equilibrium, using the flux coordinate system introduced in section
\ref{sec:coordmsd}. The determination of the region, where the wave deposits its energy, is important in Lower Hybrid Current Drive. The central
energy deposition seems to be prevented owing to the Electron Landau Damping when the wave propagates in high-temperature plasma or the parallel wave
number is large. The parallel wave number, fixed by the external antenna and evolving during the wave propagation, is a crucial parameter for the
determination of the power deposition profile. The general WKB formulation of the wave equation, as described above, is suitable for studying the
effects of the plasma equilibrium parameters on the power deposition profile, because the plasma equilibrium is automatically taken into account in
the ray equations for the phase and the field amplitude by using the flux coordinate system. Moreover, if an analytical magnetic equilibrium is known
(e.g. the Solov'ev equilibrium), the ray tracing equation system can be numerically integrated by incorporating directly the analytical magnetic
equilibrium, avoiding an interface that takes into account the numerical equilibrium (Grad-Shafranov solver) and can be a source of numerical noise
for the Runge-Kutta integrator. In order to perform this study and to test the sensitivity of the $n_{||}$ evolution (and consequently the power
deposition profile) as a function of the magnetic equilibrium, the ray tracing equations have been solved by changing the macroscopic parameters that
characterize the equilibrium, i.e. the elongation and the triangularity, in the case of FAST\cite{pizzuto:nf_fast} plasma parameters used by A.
Cardinali et al. in Ref. [{\!\!\citenum{cardinali:NF09fast_icrf}]. When calculating the absorption, for the sake of simplicity,
 only the linear ELD has been considered. It is obvious that this assumption tends to overestimate the absorption of LH waves in tokamak-like reactors,
which should be analyzed within the framework of the quasi-linear theory. Besides adopting the analytical equilibrium for studies of parametric
dependences of deposition profiles, we have compared its results with those of a realistic numerical equilibrium for the ITER Scenario 2 plasma,
characterized by the same macroscopic parameters such as elongation and triangularity. This is done in order to investigate the numerical error when
interfacing the ray tracing system to a numerical equilibrium and to demonstrate the similarity between the realistic numerical equilibrium and the
Solov'ev equilibrium, which we have adopted as analytical model for our parametric studies.

In Solov'ev equilibrium, by assuming that the pressure $p(\Psi)$ and the square of the poloidal current function $F^2(\Psi)$ are both linearly
dependent on $\Psi$, the solution of the Grad-Shafranov equation is given analytically as\cite{solovev:jetp68}
\begin{equation}
\Psi=\frac{\Psi_b}{y_b^2}\{\frac{4}{E^2}[1+(1-D)y]\frac{Z^2}{R_0^2}+y^2\}\;\;,
\end{equation}
where $y=\frac{R^2}{R^2_0}-1$. The subscript ``$b$'' and ``$0$'' above and in the following parameters such as $n_b$ and $n_0$, mean the boundary and
on-axis value respectively. The parameters $E$ and $D$ are related to the elongation $\kappa$ and triangularity $\delta$.  As a result, a Solov'ev
equilibrium is described by 6 parameters, i.e. the on-axis major radius $R_0$, $y_b$, which is the value of $R^2/R^2_0-1$ at the boundary, the
boundary poloidal flux function $\Psi_b$, the on-axis poloidal current function $F_0$, $E$ and $D$; summarizing, the parameter list is $(R_0, y_b,
\Psi_b, F_0, E, D)$. On the other hand, the input macroscopic parameters are the major radius $R_c$, which describes the center position of the
plasma in the mid-plane, the minor radius $a$, the on-axis magnetic field $B_0$, the on-axis safety factor $q_0$, the elongation $\kappa$ and the
triangularity $\delta$, i.e. $(R_c, a, B_0, q_0, \kappa, \delta)$. In the following calculation, we use the FAST parameters\cite{pizzuto:nf_fast},
i.e. $R_c$=1.82 m, $a$=0.64 m, $q(r=a)=4\sim5$, density $n_0$=2$\times10^{14}$ cm$^{-3}$, $n_b$=5$\times10^{13}$ cm$^{-3}$, temperature
$T_{e0}=T_{i0}$=8.5 KeV, lower hybrid frequency $f$=5 Ghz and the parallel wave number at the antenna $n_{\parallel A}$=2. $B_0$ is derived from the
equilibrium, holding $T_{e0}=T_{i0}$ and $n_{e0}=n_{i0}$ constant for different shaped equilibrium. The values of $\kappa$ and $\delta$ are given as
parameters to investigate shaping effects. The fitted density profile is
\begin{displaymath}
n(r) = \left\{ \begin{array}{ll}
 n_0\;\;, & \textrm{if }0\le r\le r_T\;\;,\\
 n_0\left[1-\alpha_n(r-r_T)^2\right]\;\;, &\textrm{if } r_T\le r\le 1\;\;,
  \end{array} \right.
\end{displaymath}
which becomes the usual parabolic profile in the whole $r$ range when $r_T=0$.

The first series of ray tracing calculation aimed at comparing the evolution of parallel wave number and amplitude considering three different
elongated plasmas, $\kappa=1$, $1.5$, $2$, at fixed triangularity $\delta=0$, for three different injection angles, a) equatorial, b) top launching
($\pi/2$) and c) bottom launching ($-\pi/2$). Numerical results are illustrated in the sequence of figures
(\ref{fig:shapen_parE})--(\ref{fig:shapeamp2dE}). First, figure (\ref{fig:shapen_parE}) demonstrates the evolution of the parallel wave number along
the ray trajectory and the absorption. The absorption's contribution to the amplitude is described by
\begin{equation}A_{abs}=\exp\{\int^r_{r_A}d\tau\gamma_{es}\}\;\;,\end{equation} where
\begin{equation}\gamma_{es}=-\frac{2\sqrt{\pi}\omega_{pe}^2c}{v_{eth}^3n_{||}}\exp\left\{-\left(\frac{c}{n_{||}v_{the}}\right)^2\right\}\end{equation}
corresponding to the given dispersion relation $D_0=0$, defined above. The results in figure (\ref{fig:shapen_parE}) show that the elongation causes
a downshift of $n_{||}$ for the wave launched from $\theta_A=0$ and thus leads to more central absorption. However, the effects is the opposite for
waves launched from $\pm\pi/2$, where the $n_{||}$ shifts upwards and the absorption layer moves peripherally because of elongation. Here, we notice
that the absorption of the wave depends on both $T_e$ and $n_{||}$. If the linear absorption of the wave is dominated by the local value of the
electron temperature, then the variation induced by the equilibrium quantities on the parallel wave number is ineffective on the localization of the
absorption layer\cite{cardinali:pop93shaping_effect}. Here, however, since the $T_e(\Psi)$ profile is relatively flat, the variation of $n_{||}$
might play a significant role in the localization of the absorption layer. At the next order in the WKB expansion, figure (\ref{fig:shapeampE}) shows
the evolution of the wave owing to the focusing/defocusing effects. We can deduce that the reflection position and the focal point, where the
amplitude peaks, are not the same point, as in the case of the wave propagation in the symmetric cylindrical geometry where
\begin{equation}A(r)=A(r_A)\sqrt{\frac{k_{rA}r_A}{k_r r}}\;\;.\end{equation} During a single pass, the amplitude of the wave launched from
$\theta_A=\pi/2$ peaks before reaching the reflection point and the peak position moves towards the boundary when increasing the elongation. To
investigate the geometry's effect on the focusing and de-focusing of the rays, a 2D plot of the the field amplitude in the poloidal section is shown
in figure (\ref{fig:shapeamp2dE}). The amplitude increases/decreases where the rays converges/diverges. The geometry of the equilibrium in fact could
change the optical properties of the plasma that in this case is acting like a lens.

\ifthenelse{\equal{\plotfigure}{false}}{}{
\begin{figure*}\centering
\includegraphics[width=0.85\textwidth]{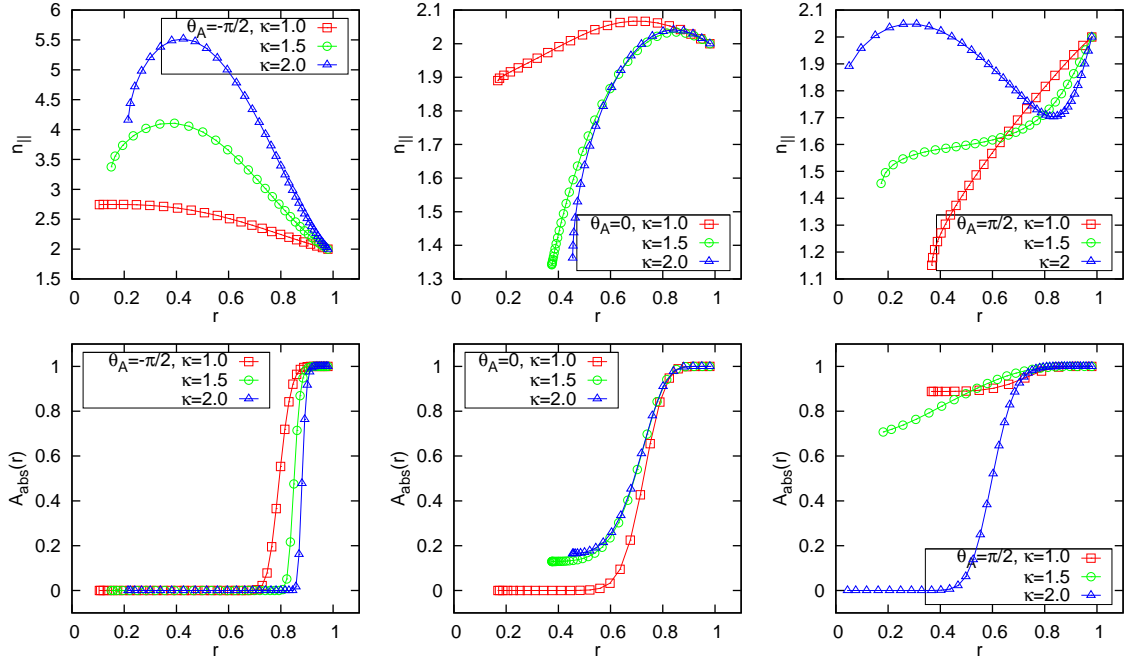}
\caption{The effects of elongation on $n_{||}$ and the LH wave absorption.}\label{fig:shapen_parE}
\end{figure*}
\begin{figure*}\centering
\includegraphics[width=0.85\textwidth]{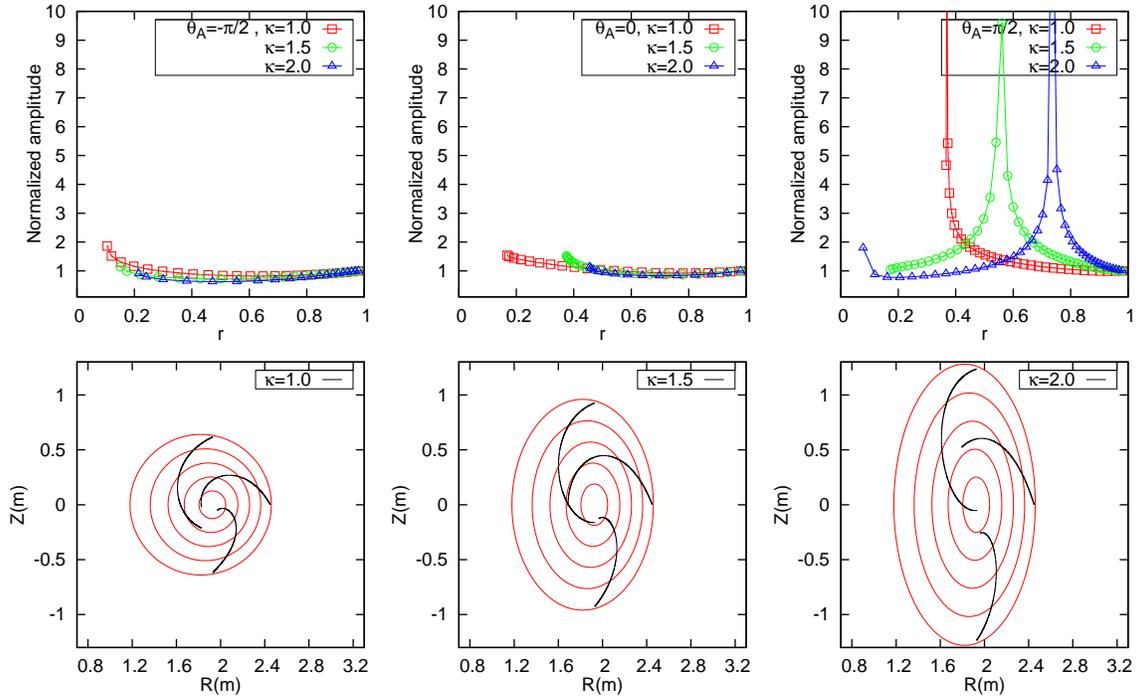}
\caption{The effects of elongation on the LH wave amplitude.}\label{fig:shapeampE}
\end{figure*}
\begin{figure*}\centering
\includegraphics[width=0.95\textwidth]{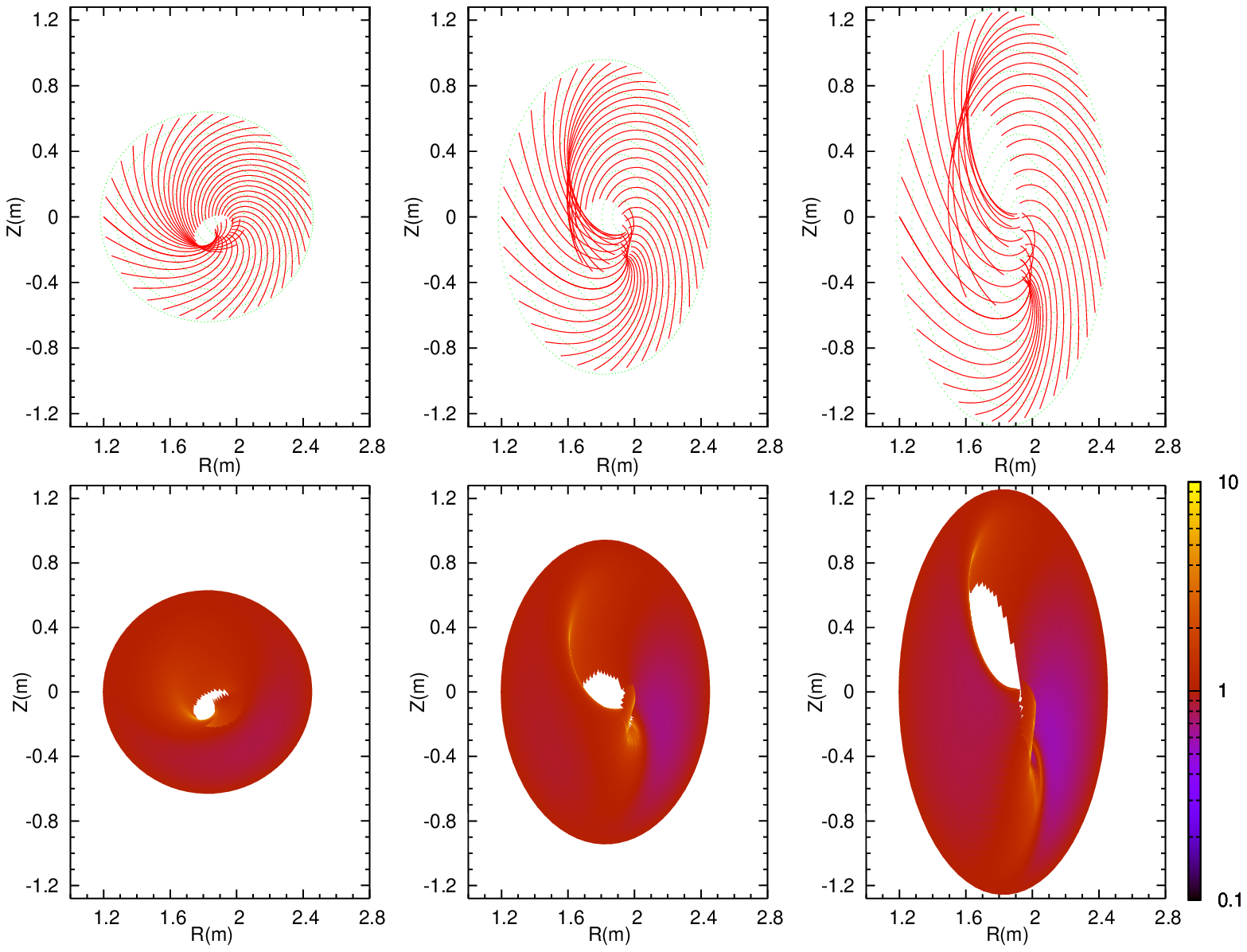}
\caption{The 2D plot of the LH wave amplitude in equilibria with different elongation.}\label{fig:shapeamp2dE}
\end{figure*} }

A second comparison regards the evolution of the parallel wave-number and amplitude by considering four different triangularity values, $\delta=0$,
$0.1$, $0.2$, $0.4$, at fixed elongation $\kappa=1$, for $\theta_A=0,\pm\pi/2$. Figure (\ref{fig:shapen_parD}) shows that higher triangularity brings
an upshift of $n_{||}$ for $\theta_A=0$ and more peripheral absorption, while the effects are the opposite for waves launched from $\pi/2$. The
effect of the triangularity on absorption of the wave launched from $\theta_A=-\pi/2$ is not manifest, since $n_{||}$ increases to a level where most
of the wave is absorbed in a thin peripheral layer near the same position and there is no evidence of how the triangularity influences the evolution
of $n_{||}$. Figure (\ref{fig:shapeampD}) and (\ref{fig:shapeamp2dD}) give the 1D and 2D plot of the amplitude, which shows that the wave launched
from $\theta_A=\pi/2$ is more affected by triangularity than those launched from $\theta_A=0, -\pi/2$. Since the triangularity makes the propagation
more peripheral for $\theta_A=\pi/2$ in this case, the focal point also moves in the same direction.

\ifthenelse{\equal{\plotfigure}{false}}{}{
\begin{figure*}\centering
\includegraphics[width=0.85\textwidth]{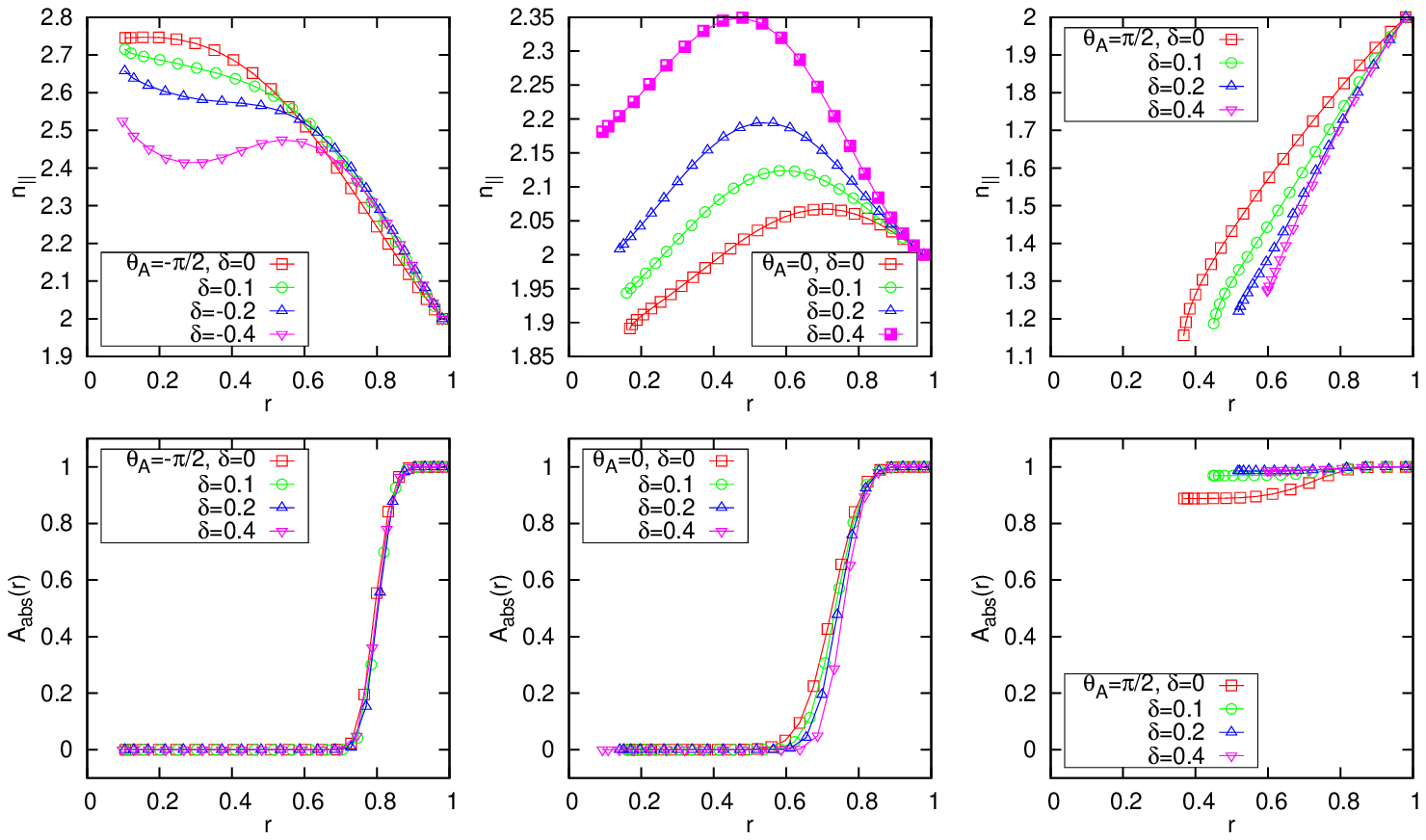}
\caption{The effects of triangularity on $n_{||}$ and the LH wave absorption.}\label{fig:shapen_parD}
\end{figure*}
\begin{figure*}\centering
\includegraphics[width=0.85\textwidth]{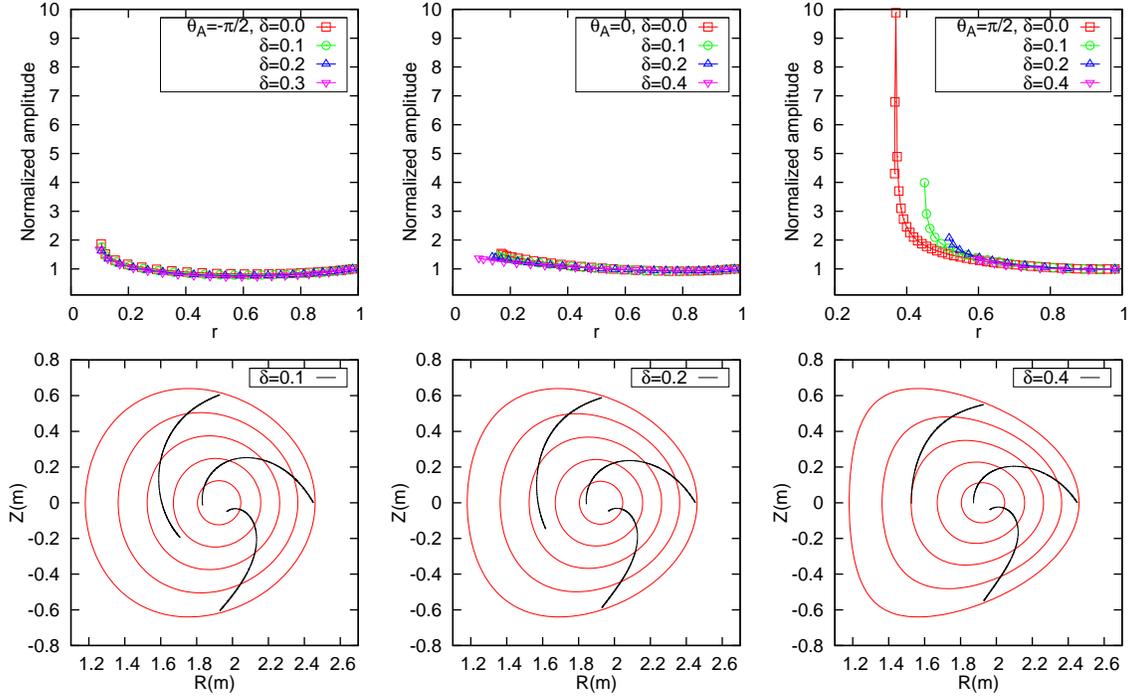}
\caption{The effects of triangularity on the LH wave amplitude.}\label{fig:shapeampD}
\end{figure*}
\begin{figure*}\centering
\includegraphics[width=0.95\textwidth]{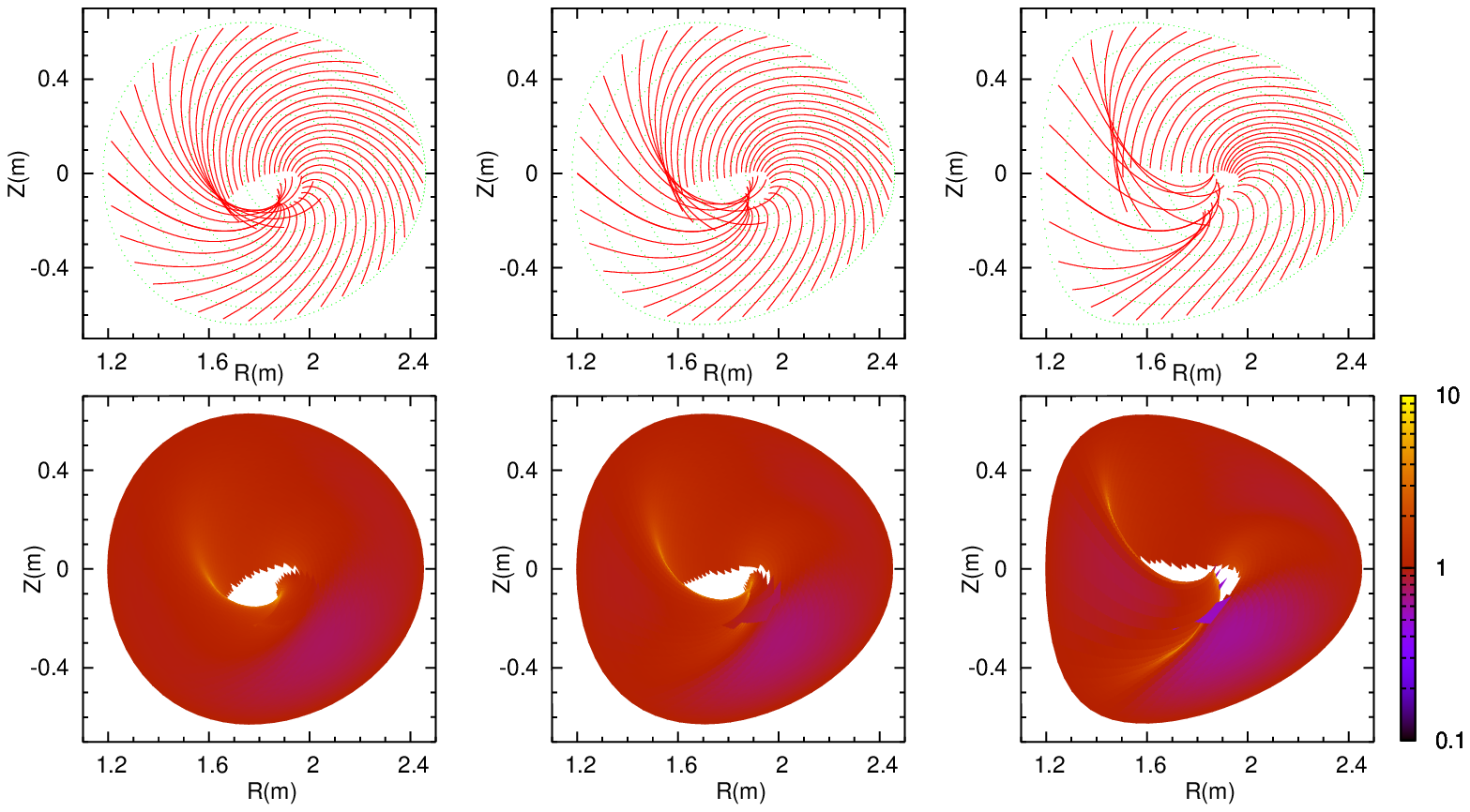}
\caption{The 2D plot of the LH wave amplitude in equilibria with different triangularity.}\label{fig:shapeamp2dD}
\end{figure*} }

Finally, figure (\ref{fig:cmpraytrace}) shows a comparison of LHW propagation in the case of ITER Scenario 2, when using a numerical equilibrium file
and a Solov'ev equilibrium characterized by the same macroscopic parameters $(R_c, a, B_0, q_0, \kappa, \delta)$. In this case, an accurate
evaluation of the error in integrating the ray tracing equations is required and compared with the analytical equilibrium case. To verify the
numerical accuracy using the numerical equilibrium, we generate the standard EQDSK file for the numerical Solov'ev equilibrium
 using the same parameters $(R_c, a, B_0, q_0, \kappa, \delta)$ and calculate the ray trajectory, parallel wave number and the relative error.
The results from the numerical and analytical Solov'ev equilibrium are in good agreement. Besides that, the propagation is calculated with the
numerical ITER equilibrium. In figure (\ref{fig:cmpraytrace}), we can see that the fitted Solov'ev equilibrium produces similar results as the ITER
numerical equilibrium. However, some differences can be observed because the fitted Solov'ev equilibrium and the ITER numerical one are not
rigorously equivalent, owing to the up-down asymmetry of the ITER numerical equilibrium, which is, obviously, more realistic than the analytical
Solov'ev equilibrium. In addition, the Solov'ev equilibrium is characterized by $p(\Psi)$ and $F^2(\Psi)$ that vary linearly with $\Psi$. This is not
the case for a real ITER equilibrium. However, the close resemblance of the results for both cases shows that it is reasonable to use a fitted
Solov'ev equilibrium for a preliminary parametric investigation of the shaping effect, to be complemented by more detailed investigations adopting a
more realistic equilibrium, after the interesting parameter range is identified. In addition, assuming the analytical equilibrium reduces the
numerical error in the Runge-Kutta integrator with respect to the numerical equilibrium, as well the integration time, avoiding the interpolation of
the flux function at each step of the calculation. In figure (\ref{fig:cmpraytrace}), the relative error along the ray trajectory is plotted in the
case of the analytical/numerical Solov'ev equilibrium and the ITER numerical equilibrium. The analytical Solov'ev reduces the numerical error by a
factor of $\sim10^2$.

\ifthenelse{\equal{\plotfigure}{false}}{}{
\begin{figure*}\centering
\includegraphics[width=0.75\textwidth]{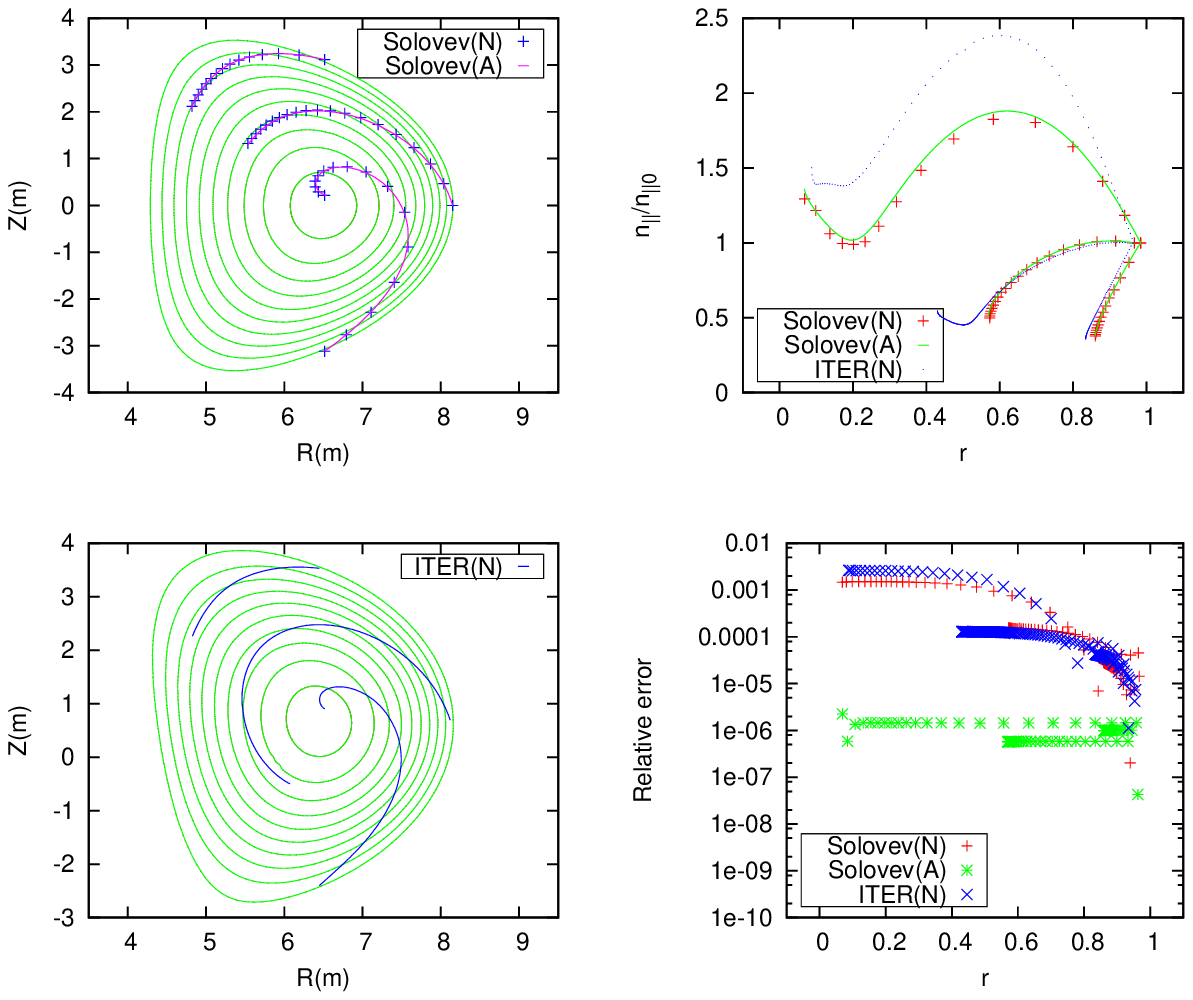}
\caption{The ray trajectories, $n_{||}$ evolution and relative numerical error of the wave propagation in numerical ITER equilibrium (ITER(N)),
analytical Solov'ev equilibrium (Solov'ev(A)) and numerical Solov'ev equilibrium (Solov'ev(N)).}\label{fig:cmpraytrace}
\end{figure*} }

This ray-tracing analysis in a general magnetic-field equilibrium shows the dependence of the propagation and absorption of LH waves on the main
parameters of the plasma equilibrium, namely, elongation and triangularity, as well as on the poloidal angle of injection (equatorial, top, bottom).
The results of the analysis can be summarized as follows. In the case of a flat density profile, which is relevant for ITER operations, it has been
shown that elongation causes a down-shift of $n_{||}$ and thus more central absorption of the wave energy for the equatorial injection; while, for
top/bottom injection, $n_{||}$ shifts upward and the deposition layer is more peripheral. On the contrary, triangularity leads to up-shift of
$n_{||}$ and more peripheral energy deposition for equatorial injection; while, for top injection, $n_{||}$ shifts downward and the deposition layer
is more central. Here, we also remind the conclusion by A. Cardinali in Ref. [\!\!\citenum{cardinali:pop93shaping_effect}] that the variation induced
by the equilibrium quantities on the parallel wave number might be ineffective on the localization of the absorption layer if the linear absorption
of the wave is dominated by the local value of the electron temperature. For example, if the temperature has a large gradient near the boundary, then
most of the energy would be deposited in a narrow layer, where the temperature is high enough for the electron Landau damping, regardless the values
of elongation and triangularity. The shaping effect on $n_{||}$ and the position of energy deposition layer is summarized in Table
\ref{table:shaping_effect}, where we can see that the elongation and triangularity tend to have opposite effects. At the next order in the WKB
expansion, the amplitude is calculated and the 2D electric field is reconstructed, demonstrating the focusing/defocusing of the rays when propagating
in a shaped equilibrium. The comparison of the ray trajectory and evolution of $n_{||}$ in analytical and numerical equilibria shows the validity of
the investigation using the Solov'ev equilibrium model and also reveals advantages and drawbacks of using the fitted analytical equilibrium.\\

\begin{table}
\centering
\begin{tabular}{cccc}
\hline\hline
Geometry & $\theta_A$ & $n_{||}$ shift & Absorption layer \\
\hline \; & $-\pi/2$ & up-shift & peripheral\\ \cline{2-4} $\kappa$ & 0 & down-shift & central\\ \cline{2-4}
\; & $\pi/2$ &\; down then up shift \;& \;mainly peripheral\;\\
\hline \; & $-\pi/2$ & up/down shift & not evident\\ \cline{2-4} $\delta$ & 0 & up-shift & peripheral\\ \cline{2-4}
\; & $\pi/2$ & down-shift & central\\
\hline\hline
\end{tabular}
\caption{The effects of elongation and triangularity on the evolution of $n_{||}$ and the position of the enegy deposition
layer. $\theta_A$, $\kappa$, $\delta$ stand for the launching angle, elongation and triangularity, respectively.}\label{table:shaping_effect}
\end{table}

%%%%%%%%%%%%%%%%%%%% setction 3B ITG
\subsection{Electrostatic ITG propagation in tokamaks and eigenmode formation}
\label{sec:ITG} As an application of the mixed WKB--full-wave approach, we discuss the case of electrostatic  ITG propagation in tokamaks and
eigenmode formation, using the ITG wave equation in the fluid limit for a low-pressure tokamak plasma, with shifted circular magnetic flux surfaces,
as reported in Ref. [\!\!\!\:\!\: \citenum{romanelli93}]. Using the MSD approach and spatial scale separation between equilibrium profiles and radial
mode envelope (see Section \ref{sec:coordmsd}), we decompose the perturbed potential $\Phi(r,\theta,\zeta,t)$ in the Ballooning Formalism
representation
\begin{eqnarray}\label{eq:msd_itg}
&\Phi(r,\theta,\zeta,t) & =  e^{in\zeta}\phi(r,\theta,t)\;\;,\\ \nonumber &\phi(r,\theta,t) & =  2\pi A(r,t)\sum_p \delta\phi(\theta+2p\pi)
e^{-inq(\theta+2p\pi)}\;\;,
\end{eqnarray}
where $\delta\phi(\theta)$ is the parallel mode structure in the mapping space (see Section \ref{sec:coordmsd}, Appendix \ref{app:msd} and
\ref{app:ballooning}). This equation is equivalent to equation (\ref{eq:ballooning_cleb}) and
 $\delta\phi(\eta)$ corresponds to $\bar F(r,\eta)$ there. Writing $A(r,t)$ in the eikonal representation
\begin{eqnarray}
A(r,t)=e^{-i\omega t+i\int nq'\theta_kdr}\;\;,
\end{eqnarray}
we can get the equation for the parallel structure in [\!\!\citenum{romanelli93}]
 \begin{eqnarray}
&\frac{\omega_{ti}^2}{\omega^2} \frac{\partial^2}{\partial \theta^2} \delta \phi + \Big\{ \frac{1/\tau + \omega_{*ni}/\omega}{1 -
\omega_{*pi}/\omega} + (k_\theta^2 \rho_i^2) \left[ 1 + s^2 (\theta - \theta_{k0})^2 \right] &  \nonumber
\\ & - \frac{\omega_D}{\omega} \left[ \cos \theta + s (\theta - \theta_{k0}) \sin \theta \right]\Big\} \delta \phi \,=\, 0 &\;\; . \label{eq:ITG}
\end{eqnarray}
Here $\omega_{ti}=v_{ti}/(qR_0)$, $v_{ti}=(T_i/m_i)^{1/2}$ is the ion thermal speed, $R_0$ is the tokamak major radius, $\tau=T_e/T_i$, $\omega_{*ni}
= (T_i c/e B)(
$$\bf k \times b$$)\cdot$$\bf(\nabla$$ n_i)/n_i$, $\omega_{*Ti} = (T_i c/e B)($$\bf k \times b$$)\cdot$$\bf (\nabla$$ T_i)/T_i$,
$\omega_{*pi}=\omega_{*ni}+\omega_{*Ti}$, $n_i$ is the thermal ion particle density, for which we have assumed unit electric charge $e$, $T_i$ is
their temperature in energy units, $\omega_{ci}$ is the thermal ion cyclotron frequency, $\rho_i=v_{ti}/\omega_{ci}$ is their Larmor radius, $\bf k =
-i \bf \nabla$ is the wave-vector, $k_\theta \simeq (-nq)/r$ is its poloidal component, $\bf b$ is the unit vector aligned with the equilibrium
magnetic field, $s=(r/q)(dq/dr)$ is the magnetic shear and $\omega_D=-2k_\theta \rho_i v_{ti}/R_0$ is the thermal ion magnetic drift frequency. The
subscript $0$ in $\theta_{k0}$ denotes the lowest (zeroth) order term in the asymptotic series expansion for $\theta_k$.

Equation (\ref{eq:ITG}) readily demonstrates the validity of  the mixed WKB-full-wave approach, since the perpendicular components of the group
velocity are introduced by the Finite Larmor Radius (FLR) effect and thermal ion magnetic drift. Thus, the perpendicular group velocity is much
smaller than the parallel group velocity. As a result, the ITG wave propagates mainly along the field line and circulates in the flux surface several
times without significant propagation in the perpendicular direction. The corresponding parallel mode structure can be obtained by solving equation
(\ref{eq:ITG}), with proper boundary conditions, i.e. outgoing wave or decaying boundary conditions. The corresponding local eigenvalue $\omega$ is
derived from the local dispersion relation
\begin{equation}\label{eq:D0=0itg}D_0(r,\theta_{k0},\omega)A(r,t)=0\end{equation} as a function of $\theta_{k0}$, $\omega$ and the parameters characterizing the local plasma
equilibrium. In other words, $\omega=\omega(r,\theta_{k0})$ from equation (\ref{eq:D0=0itg}). The WKB solution of the amplitude (radial envelope) can
be written as
\begin{eqnarray}\label{eq:amp_decomposition}
A(r,t)=A_0(r)\bar A(r,t)e^{-i\omega t}\;\;,
\end{eqnarray}
where $A_0(r)=e^{i\int nq'\theta_{k0}dr}$ is the zeroth order solution and $\bar A(r,t)$ is the higher order correction, which contains variation on
the slow spatio-temporal scales only.

In order to obtain the higher order correction for the radial envelope, we reconstruct the governing differential equation from
$D_0(r,\theta_{k0},\omega)$, by substitution of $\omega\Rightarrow i\partial_t$ and $\theta_{k0}\Rightarrow(-i/nq')\partial_r$, and expanding it
locally along the characteristic of the wave-packet propagation in phase space as\cite{chen:prl2004_zf,zonca:pop04NL_interplay}
\begin{eqnarray}\label{eq:amp_differential}
&&\frac{\partial D_0}{\partial\omega}\left(i\frac{\partial}{\partial t}-\omega\right)A+\frac{\partial
D_0}{\partial\theta_{k0}}\left(-\frac{i}{nq'}\frac{\partial}{\partial r}-\theta_{k0}\right)A
\\ \nonumber
&&+\frac{1}{2}\frac{\partial^2D_0}{\partial\theta_{k0}^2}\left[\left(-\frac{i}{nq'}\frac{\partial}{\partial
r}-\theta_{k0}\right)^2A-\frac{i}{nq'}\frac{\partial\theta_{k0}}{\partial r}A\right]=S(r,t)\;\;,
\end{eqnarray}
where \begin{eqnarray} \left(-\frac{i}{nq'}\frac{\partial}{\partial r}-\theta_{k0}\right)^2A & = & \left(-\frac{i}{nq'}\frac{\partial}{\partial r}-\theta_{k0}\right)\\ \nonumber
&&\times\left(-\frac{i}{nq'}\frac{\partial}{\partial r}A-\theta_{k0}A\right)\nonumber\;\;,\end{eqnarray} $\partial_t$ and $\partial_r$
on the left hand side formally act on quantities that follow and the source $S(r,t)$ on the right hand side can represent the drive due to an
``internal/external'' antenna or nonlinear interactions. This is equivalent to the differential equation for the nonlinear system in
[\!\!\!\:\citenum{chen:prl2004_zf,zonca:pop04NL_interplay}], where the source term is due to the drift wave-zonal flow nonlinear interaction. After
substituting equation (\ref{eq:amp_decomposition}) into equation (\ref{eq:amp_differential}), and assuming that the trivial fast time scale variation
$\approx e^{-i\omega t}$ is isolated from the source
\[S=\bar Se^{-i\omega t}\;\;,\] we obtain the equation for the higher order correction $\bar A$
\begin{eqnarray}\label{eq:amp_differential_barA}
&&i\frac{\partial D_0}{\partial\omega}\frac{\partial}{\partial t}\bar A+\frac{\partial D_0}{\partial\theta_{k0}}\frac{1}{inq'}\frac{\partial}{\partial
r}\bar A+\frac{1}{2}\frac{\partial^2D_0}{\partial\theta_{k0}^2} \\ \nonumber
&&\times\left[\frac{1}{inq'}\frac{\partial}{\partial
r}\left(\frac{1}{inq'}\frac{\partial}{\partial r}\bar A\right)+\frac{\bar A}{inq'}\frac{\partial\theta_{k0}}{\partial r}\right]=\bar S(r,t)/A_0\;\;.
\end{eqnarray}
In the square bracket, the first term is much smaller than the second term in the region where the WKB approach applies. Far from the turning points,
ignoring the first term in the square bracket and using the ray tracing equations
\begin{eqnarray}
&&\frac{dr}{d\tau} = -\frac{1}{nq'}\frac{\partial D_0}{\partial\theta_{k0}}\;\;,\\ \nonumber &&\frac{d\theta_{k0}}{d\tau} =
\frac{1}{nq'}\frac{\partial D_0}{\partial r}\;\;,\\ \nonumber &&\frac{d t}{d\tau} = \frac{\partial D_0}{\partial\omega}\;\;,
\end{eqnarray}
we can obtain the first order correction for $A$ as
\begin{eqnarray}\label{eq:amp_int_along_tau}
A(r,t)&=&A_0(r)A_1(r)\tilde A(r,t)e^{-i\omega t}\\ \nonumber
&&\times\left[A_{S}|_{\tau=0}+\int_0^\tau d\tau\bar S/(iA_0A_1)\right]\;\;,
\end{eqnarray}
where \begin{equation}A_1=\frac{\left(A_1\sqrt{\partial D_0/\partial\theta_{k0}}\right)|_{\tau=0}}{\sqrt{\partial
D_0/\partial\theta_{k0}}}\end{equation} and the factor $\partial D_0/\partial\theta_{k0}$ reflects the focusing/defocusing effects, which is
represented by the Jacobian of the ray coordinates $J_R$ in the 2D WKB method for the lower hybrid wave propagation (see Section \ref{sec:lhcold}).
Meanwhile, $\tilde A(r,t)$ represents the even higher order corrections while the source's contribution is the integral along the characteristics of
the wave-packet propagation. When $S=0$, the wave-packet exhibits only the free streaming propagation, described by the phase variation,
focusing/defocusing effects and higher order corrections. Near the reflection point where $\partial D/\partial\theta_{k0}\rightarrow0$ and
$\theta_{k0}\rightarrow\theta_{k0T}$, WKB breaks down and the local full wave solution can be obtained from the local expansion of
$D_0(r,\theta_{k0},\omega)$ at $r_T$
\begin{eqnarray}\label{eq:amp_differential_airy}
&&\left[i\frac{\partial D_0}{\partial\omega}|_{r_T,\theta_{k0T}}\frac{\partial}{\partial t}+(r-r_T)\frac{\partial D_0}{\partial
r}|_{r_T,\theta_{k0T}}\right. \\ &&\nonumber \left.-\frac{1}{2(nq')^2}\frac{\partial^2D_0}{\partial\theta_{k0}^2}|_{r_T,\theta_{k0T}}\frac{\partial^2
}{\partial r^2}\right](A_0\bar A)=\bar S(r,t)\;\;,
\end{eqnarray}
where $r_T$ is the turning point position where $\theta_{k0}=\theta_{k0T}$ and, with $\bar S=0$, it becomes equation (23) in
[\!\!\citenum{cardinali03}]. Then, WKB solution in its validity region and the local full wave solution at the reflection point are matched
asymptotically to give the solution in the whole radial range. After normalization and Laplace transform in time of $(A_0\bar A)$, the local full
wave equation reduces to the Airy function equation from where one can derive the connection formulae \mbox{[\!\!\citenum{cardinali03}]}. In the case
of an isolated mode with exponentially decaying boundary conditions outside its local support, we readily get the phase shift of $\pm\pi/2$ between
the incident wave and the reflected wave at the reflection layer. In the more general case, the wave-packet can still propagate and tunnel through
the cutoff layer, with given reflection and transmission coefficients, and eventually reach the boundary, henceforth bouncing back and forth.
Application of global boundary conditions allows computing reflection and transmission coefficients at all turning points. After long enough time,
the eigenmode structure is eventually generated if it exists.

The envelope tracing method discussed here, is different from the eigenvalue approach in Refs. [\!\citenum{romanelli93,taylor93}], where the radial
eigenfunction is obtained using the decaying boundary condition at $r\rightarrow\pm\infty$. In the envelope tracing approach, the wave-packet
tunneling and reflection at the turning points can be taken into account and be described with the corresponding transmission and reflection
coefficients to be determined from global boundary conditions. The wave-packet, on the one hand, bounces forth and back in the propagation region
and, on the other hand, can reach other regions of propagation or cutoff. These physics can be implemented systematically in a numerical solution
scheme, while a direct eigenvalue approach requires the computation of global phase integral quantization condition on complex phase-space contours
that are dependent on the eigenvalue itself. Thus, the initial value approach is easier to perform numerically than the eigenvalue approach, except
for the simple case of an isolated mode, where the two methods are essentially equivalent and straightforward. After the parallel structure and
radial envelope propagation are obtained, the time varying 2D field $\phi(r,\theta)$ can be reconstructed from the solution of equation
(\ref{eq:ITG}), $\delta\phi(\theta)$, and the amplitude $A(r,t)$, using equation (\ref{eq:msd_itg}).

The 2D structure of ITG mode is important for understanding anomalous transport. The radial structure of the ITG mode has been investigated in
previous literature [\!\!\citenum{romanelli93,connor93,taylor93,taylor96}]. Here, we go further and analyze the time varying {2D ITG mode structure
by means of an initial value approach, which also makes it possible to compare our findings with other 2D ITG mode solvers. The shaping effects from
the equilibrium geometry can also be readily investigated adopting general coordinates, as shown in the case of lower hybrid wave propagation
discussed in Section \ref{sec:lhcold}.

In order to calculate the $\delta\phi$ numerically, we reduce equation (\ref{eq:ITG}) to the dimensionless form
%\begin{widetext}
\begin{eqnarray}
\frac{\partial^2}{\partial \theta^2} \delta \phi &+& \Omega^2\left\{ \frac{\Omega/\tau + \Omega_{*ni}}{\Omega - \Omega_{*pi}} + (k_\theta^2 \rho_i^2)
\left[ 1 + s^2 (\theta - \theta_{k0})^2 \right] \right\} \delta \phi \nonumber
\\ & & - \Omega_D\Omega \left[ \cos \theta + s (\theta - \theta_{k0}) \sin \theta \right] \delta \phi \,=\, 0 \;\; , \label{eq:ITG_norm}
\end{eqnarray}
%\end{widetext}
where $\Omega=\omega/\omega_{ti}$, $\Omega_D=\omega_D/\omega_{ti}=-2k_\theta\rho_i q$, $\Omega_{*ni}=\omega_{*ni}/\omega_{ti}=-k_\theta\rho_i q
R_0/L_n$ and $\Omega_{*pi}=\omega_{*ni}/\omega_{ti}=-k_\theta\rho_i q R_0/L_p$. We consider typical parameters at the reference radial position $r_0$
to be $s=1$, $q=2$, $k_\theta\rho_i=0.3$, $\tau=1$, $R_0/L_T=3$, $R_0/L_n=0$. The ion temperature gradient profile is assumed in the form
$g(x)=4(R_0/L_T)(e^{x}+e^{-x})^{-2}$, where $x$ is the normalized shifted radial coordinates $x=(r-r_0)/\Delta r$. In the following, we choose
$r_0=0.5a$, $\Delta r=0.2a$, while the toroidal mode number is assumed to be $n=38$ from $k_\theta\rho_i=nq\rho_i/r\approx 2nq\rho_i/a$, where
$k_\theta\rho_i\approx 0.3$ and $a/\rho_i\approx500$. The safety factor gradient $q'$ can be estimated as $q'=sq/r\approx4/a$ and thus the eikonal
$\int{drnq'\theta_{k0}}\approx30.4\int{dx \theta_{k0}}$.

Figure (\ref{fig:parallel_structure}) gives the parallel structure in the mapping space at $x=0$. The parallel mode structure is localized and peaked
near $\theta=\theta_{k0}$ as expected. Figure (\ref{fig:omega_r}) shows the ion temperature gradient profile (left) and the local eigenvalue
variation with respect to $R_0/L_T$ (right).  The local growth rate, in the form of the imaginary part of $\Omega(r,\theta_{k0})$, increases when the
ion temperature gradient $R_0/L_T$ increases. The local growth rate is also affected by $\theta_{k0}$, since the parallel mode structure is peaked
near $\theta-\theta_{k0}\sim0$ and, thus, for $\theta_{k0}=0$ the mode is localized in the bad curvature region and is most unstable.

\ifthenelse{\equal{\plotfigure}{false}}{}{
\begin{figure*}
\includegraphics[width=0.3\textwidth]{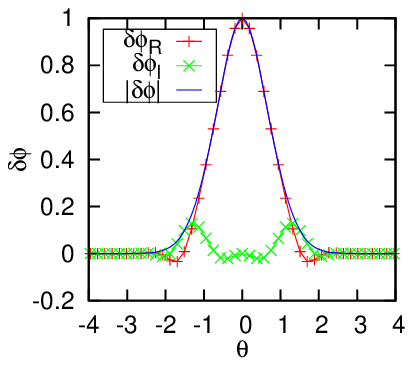}
\includegraphics[width=0.3\textwidth]{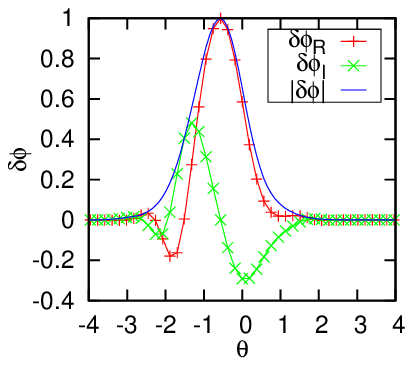}
\includegraphics[width=0.3\textwidth]{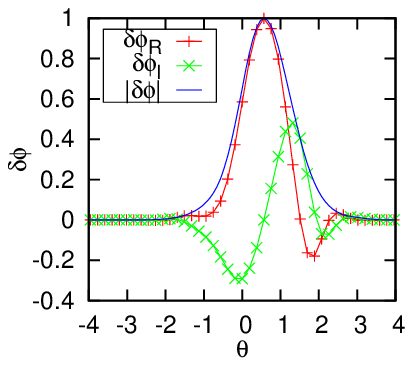}
\caption{The ITG parallel mode structure in the mapping space, with parameters  $s=1$, $q=2$, $k_\theta\rho_i=0.3$, $\tau=1$,$R_o/L_T=3$,
$R_o/L_n=0$, $n=38$ and $\theta_{k0}=0$ (left), $\theta_{k0}=-1.0$ (center), $\theta_{k0}=1.0$ (right).}\label{fig:parallel_structure}
\end{figure*}
\begin{figure*}
\includegraphics[width=0.45\textwidth]{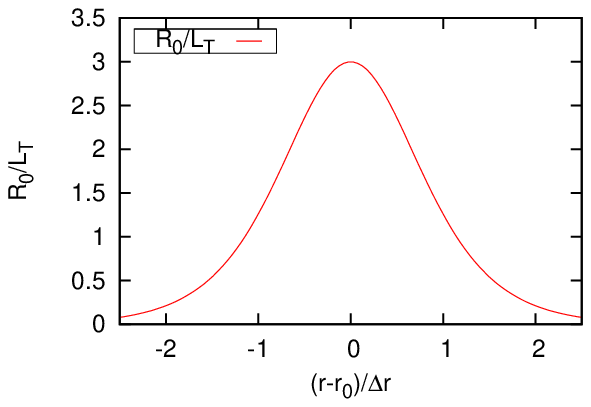}
\includegraphics[width=0.45\textwidth]{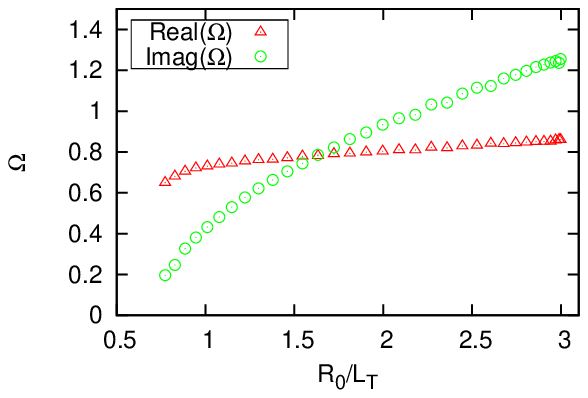}
\caption{The ion temperature gradient profile (left) and the ITG local eigenvalue $\Omega$ as a function of $R_0/L_T$ (right).}\label{fig:omega_r}
\end{figure*} }

When solving the radial envelope structure as initial value problem, the constant eigenfrequency curves generate two types of trajectories in phase
space $(\theta_{k0},r)$, classified as ``librations'' or ``rotations'' on the basis of topology arguments and investigated analytically and
numerically in [\!\!\citenum{zonca93,romanelli93,connor93,taylor93,taylor96,dewar97}]. The lowest order solution for the normalized amplitude
$A_0(r)$ in the region between turning points is shown in figure (\ref{Fig:amp_rITG}). The wave launched at the left turning point $r_{T1}$
propagates to the right turning point $r_{T2}$ (left frame in figure (\ref{Fig:amp_rITG})), where the phase is shifted by $\pi/2$, and then
propagates leftward (center frame in figure (\ref{Fig:amp_rITG})). The interference patterns are generated because the wave bounces between the
turning points, which, in this case, are located at $x=\pm0.3$ for a phase space ``libration'' (right frame in figure (\ref{Fig:amp_rITG})).  In
figure (\ref{fig:fphirhi}), the next order correction for the amplitude $A_1$ is plotted. As expected, its radial structure is flat in most part of
the propagation region around $x=0$, while near the turning points, where $\partial D_0/\partial\theta_{k0}=0$, the wave group velocity vanishes,
leading to the increase of the amplitude. For the sake of simplicity, only the wave-packet propagation region is plotted, while the exponential decay
outside the two turning points and the smooth connection of the solution of equation (\ref{eq:amp_differential_airy}) near the turning points is
implicitly assumed.

\ifthenelse{\equal{\plotfigure}{false}}{}{
\begin{figure*}
\includegraphics[width=0.3\textwidth]{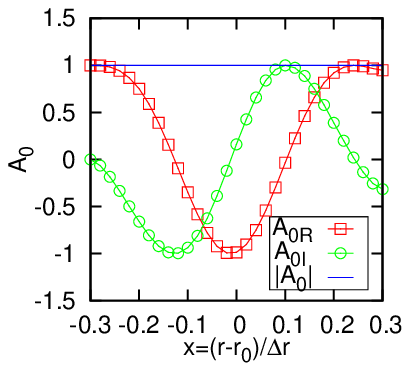}
\includegraphics[width=0.3\textwidth]{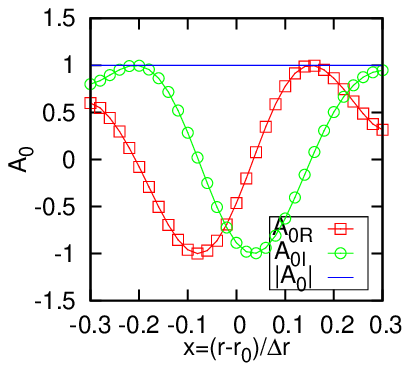}
\includegraphics[width=0.3\textwidth]{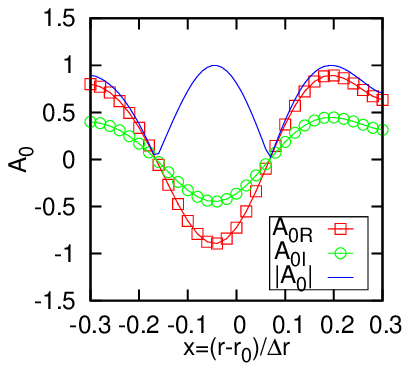}
\caption{$A_0(r)$'s evolution versus $x$. Left: propagation from the left to the right; center: propagation from the right to the left; left: the
superposition of the previous two.}\label{Fig:amp_rITG}
\end{figure*}
\begin{figure}
\includegraphics[width=0.5\textwidth]{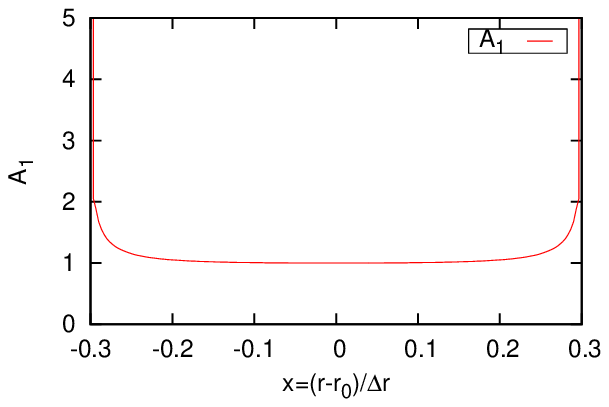}
\caption{$A_1(r)$'s evolution versus $x$.}\label{fig:fphirhi}
\end{figure} }

To illustrate the 2D time dependent ITG mode structure, we assume a point-like source, $\bar S(r,t)=\delta(r-r_A)$, where $r_A$ is the antenna
position at $r_{T1}$, $r_A=r_{T1}$. Then equation (\ref{eq:amp_int_along_tau}), with the phase shift at the reflection layer, can be rewritten as
\begin{widetext}\begin{eqnarray}
e^{i\omega t}A(r,t)=&&\exp\left\{i\int^t_0nq'\left(\theta_{k0}+\theta_{k1}\right)dr-i\sum_{l=1}^{N_T}\delta_l\pi/2\right\}\tilde A(r,t)\times\\
\nonumber
&&\left\{A_S|_{\tau=0}+\frac{1}{i}\int^{\tau(t)}_0\delta(r-r_A)\exp\left[-i\int^{\tilde\tau}_0nq'(\theta_{k0}+\theta_{k1})dr+i\sum_{l=1}^{\tilde
N_T}\delta_l\pi/2\right]d\tilde \tau\right\}
\end{eqnarray}\end{widetext}
where $N_T$ is the number of turning points the wave-packet passes by in the time interval $\tilde t\in(0,t)$ and $\delta_l=\pm1$ depending on the
phase shift of the reflected wave-packet at the turning point. Furthermore, \[\theta_{k1}=\frac{i}{2}\ln{\frac{\partial
D_0}{\partial\theta_{k0}}}\;\;.\] When propagating between the turning point pair, the wave-packet passes $r_A$ at $t=t_{A0}, t_{A1}\dots t_{AN_T}$
(see figure (\ref{fig:itg_multiref})). Assuming $A_S|_{\tau=0}=0$, $N_T$ as a positive odd integer and defining the rotation/libration number
(decreased by one) as $N=(N_T-1)/2$, we obtain
\begin{widetext}\begin{eqnarray}
e^{i\omega t}A(r,t)=&&\exp\left\{i\int^t_0nq'\left(\theta_{k0}+\theta_{k1}\right)dr-i\sum_{l=1}^{N_T}\delta_l\pi/2\right\}\tilde A(r,t)\times\\
\nonumber &&\frac{1}{i}\sum_{\tilde N_T =0}^{N_T }\exp\left\{-i\int^{t_{A\tilde N_T}}_0nq'(\theta_{k0}+\theta_{k1})dr+i\mathbb{H}(\tilde N_T-1)\sum_{l=1}^{\tilde N_T}\delta_l\pi/2\right\}\\
\nonumber =&&\frac{\tilde A(r,t)}{i}\sum_{\tilde N_T =0}^{N_T }\exp\left\{ i\int^t_{t_{A\tilde
N_T}}nq'(\theta_{k0}+\theta_{k1})dr-i\mathbb{H}(N_T-\tilde N_T-1)\sum_{l=\tilde N_T+1}^{ N_T}\delta_l\pi/2\right\}\;\;,
\end{eqnarray}\end{widetext}
where the Heaviside function $\mathbb{H}$ is used to eliminate the summation if the lower bound of its running index is larger than its upper bound.
This equation reduces to
\begin{equation}\label{eq:amp:s_n}
e^{i\omega t}A(r,t)=\hat{A}(r,t)S_N(t)\;\;,
\end{equation}
where
\begin{eqnarray}
&\hat{A}(r,t)&=\frac{\tilde{A}(r,t)}{i} \left\{\exp\left[i\int^t_{t_{AN_T}}nq'(\theta_{k0}+\theta_{k1})dr\right]\right.\\ \nonumber
&&\left.+\exp\left[i\int^{t}_{t_{A(N_T-1)}}nq'(\theta_{k0}+\theta_{k1})dr-\delta_{N_T}\pi/2\right]\right\}\;\;,
\end{eqnarray}
\begin{eqnarray}
 &S_N&=\frac{1-\hat\epsilon^{N+1}}{1-\hat\epsilon}\;\;, \;\;\hat\epsilon=\exp\{i\Phi_0\}\;\;,\;\;\\ \nonumber
&\Phi_0&=i\int_{t_{A(2m)}}^{t_{A(2m+2)}} nq'\theta_{k0}dr-\frac{i\pi(\delta_{2m+1}+\delta_{2m+2})}{2}\\ \nonumber &&=i\oint
nq'\theta_{k0}dr-\beta\pi\;\;,
\end{eqnarray}
where $\beta$ is the phase shift (Maslov index), with $\beta=1$ for librations and $\beta=0$ for rotations. Equation (\ref{eq:amp:s_n}) agrees with
the analysis in [\!\!\citenum{zonca04}] and illustrates the eigenmode formation and phase mixing. In fact, $S_N(t)$, as a function of $\Omega$ and
$N(t)$, describes interference patterns that becomes narrower around $\Phi_0=2l\pi$ for increasing $N$, corresponding the global (2D)
eigenfrequencies $\Omega_{G_l}$, with $l\in\mathbb Z$ as the radial mode number (the left frame in figure (\ref{Fig:sn})). The right frame in figure
(\ref{Fig:sn}) shows that, unless $\Omega=\Omega_{G_l}$, i.e. $\Phi_0=2l\pi$, the normalized amplitude will decay because of phase mixing when $N$
increases.  The 2D mode structure with $\Omega=(0.8582, 1.235)$ is selected as a eigenmode according to $S_N$'s time ($N$) evolution. Figure
(\ref{Fig:fphi2d}) illustrates the 2D time dependent ITG mode structure. The left top frame shows the 2D mode structure with $\Omega=(0.8582,1.235)$
and $N=0$ (for one period of the oscillation in phase space). The wave pattern and intensity remain the same for $N=8$ (the top center frame in
figure {\ref{Fig:fphi2d}}), since $\Omega=(0.8582,1.235)$ is the eigenfrequency of the 2D problem. The phase mixing of the 2D structure with
$\Omega=(0.8576,1.231)$ is illustrated in the bottom figures. The field intensity decreases to a low level due to phase mixing, while the mode
structure remains the same from $N=0$ (left bottom) to $N=8$ (center bottom) as a consequence of our present assumption of a point-like source. For a
more general source term, e.g. a broad internal source, the mode structure will also change and short wave number structures will be generated due to
phase mixing: the radial eigenmode structure is preserved asymptotically in time only for the correct global eigenvalues. The third column of figure
(\ref{Fig:fphi2d}) shows the fine structure for $\theta\in[-\pi/6,\pi/6]$, where we can recognize the radial envelope and the variation of poloidal
harmonics.

\ifthenelse{\equal{\plotfigure}{false}}{}{
\begin{figure}
\includegraphics[width=0.5\textwidth]{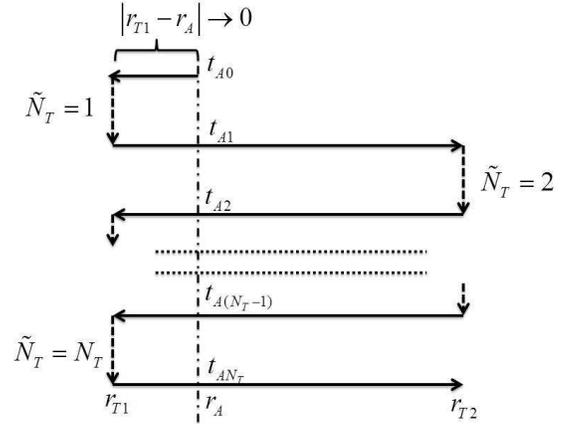}
\caption{The radial propagation of the ITG wave-packet between the turning points. The wave-packet is launched at $r_A$ and propagates between
$r_{T1}$ and $r_{T2}$.}\label{fig:itg_multiref}
\end{figure}
\begin{figure*}
\includegraphics[width=.45\textwidth]{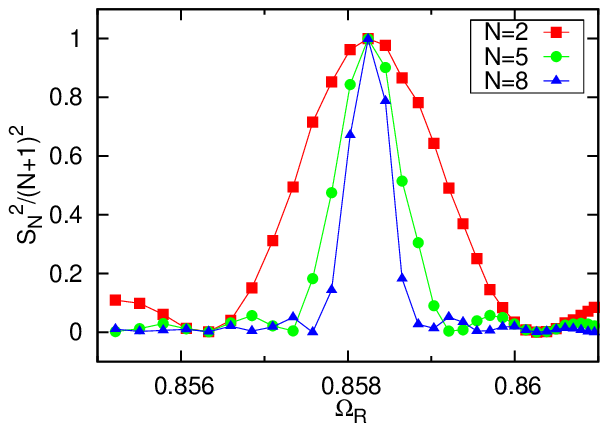}
\includegraphics[width=.45\textwidth]{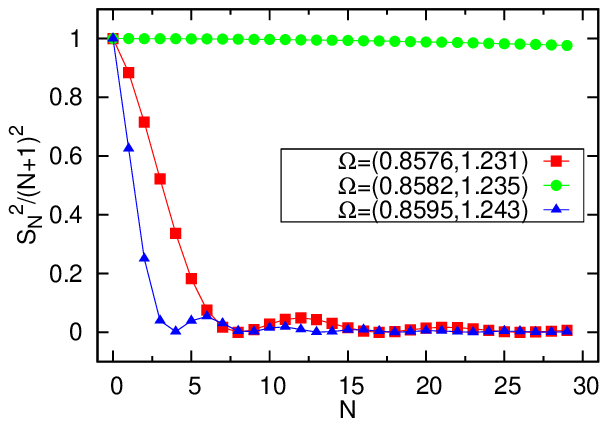}
\caption{$S_N$'s evolution versus $\Omega_R$ (left) and $N$ (right).}\label{Fig:sn}
\end{figure*}

\begin{figure*}
\includegraphics[width=.38\textwidth]{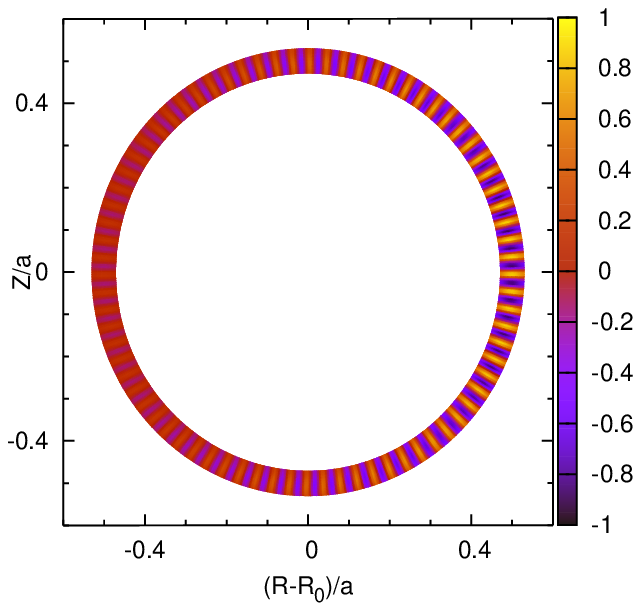}
\includegraphics[width=.59\textwidth]{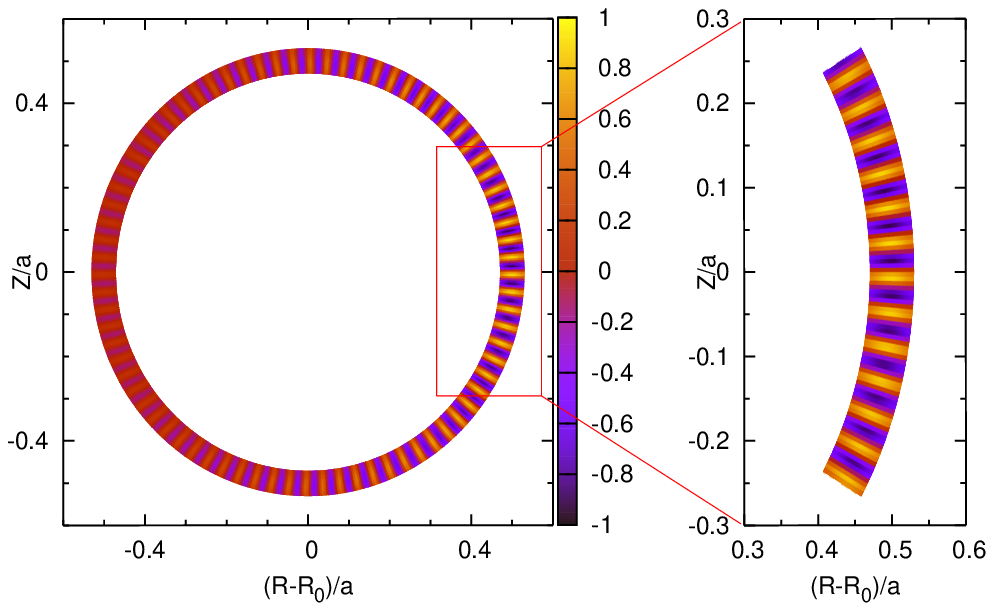}

\includegraphics[width=.38\textwidth]{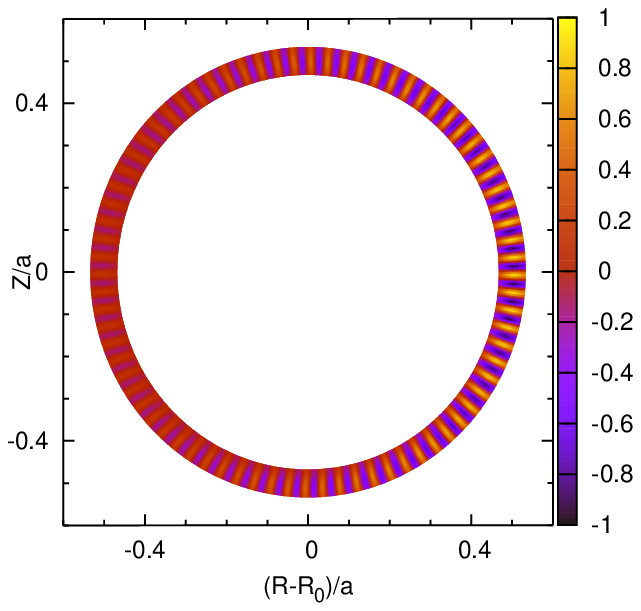}
\includegraphics[width=.59\textwidth]{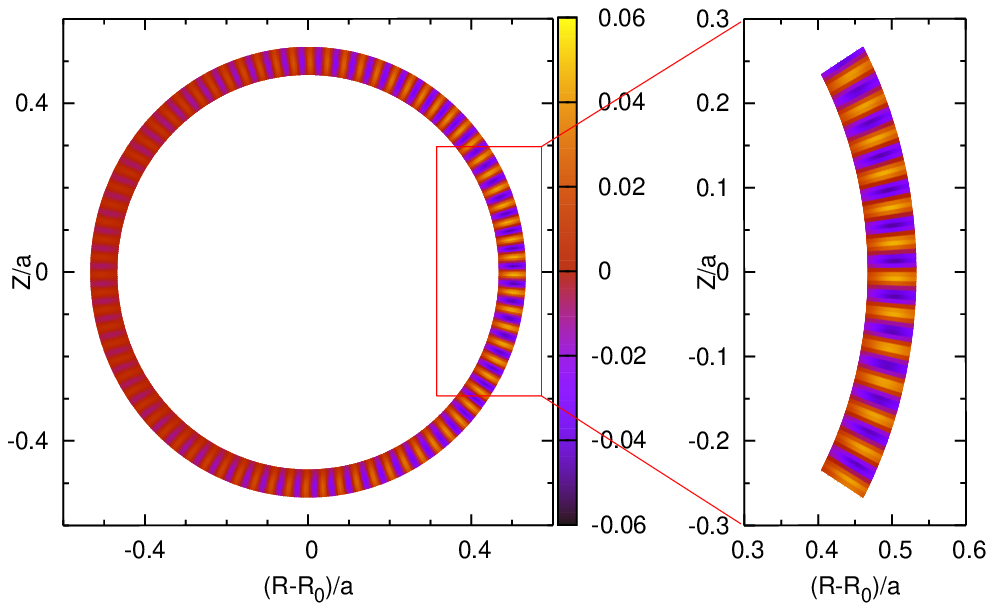}
\caption{The time evolution of the 2D field, Real$\{e^{i\omega t}\phi(r,\theta,t)\}$. Left top: $\Omega=(0.8582,1.235)$, $N=0$; center top:
$\Omega=(0.8582,1.235)$, $N=8$; left bottom: $\Omega=(0.8576,1.231)$, $N=0$; center bottom: $\Omega=(0.8576,1.231)$, $N=8$. The right figure show the
fine mode structures for $\theta\in[\pi/6, \pi/6]$ in both cases.}\label{Fig:fphi2d}
\end{figure*}}

In the example above, the mixed WKB-full-wave solution for the 2D ITG mode structure is derived within the framework of the mode structure
decomposition (MSD) method, which coincides with the Ballooning Formalism in this specific case, since spatial scale separation applies. The parallel
mode structure is obtained by solving the local eigenvalue problem on a given flux surface. The radial propagation of the envelope wave-packet has
been solved, including focusing/defocusing effects, using the WKB formulae for wave-packet propagation. At last, the time dependent 2D mode structure
is reconstructed, which illustrates the phase mixing and the 2D eigenmode formation, and also provides a picture for comparison with other 2D
numerical solvers. A more realistic investigation, with geometry effects for the 2D time dependent ITG mode structure and multiple space-time scale
analysis, including the source/sink term and nonlinear interactions, can be straightforwardly implemented in the framework of the mixed WKB-full-wave
formalism described here and assuming the general coordinate representation and MSD approach, described in Section \ref{sec:coordmsd} and Appendixes
\ref{app:msd} and \ref{app:ballooning}.

%%%% section conclusions and discussions
\section{Conclusions and discussions}
\label{sec:end} The aim of this work was providing a theoretical framework for investigating electrostatic wave-packet propagation in tokamak plasmas
with general geometry. Different techniques have been discussed, ranging from the 2D WKB approach to the mode structure decomposition (MSD) method
and the mixed WKB-full-wave analysis, which can be used for investigating RF wave propagation and absorption as well as the the time evolving
structures of MHD fluctuations and drift waves.

As application, we adopted the 2D WKB description to investigate equilibrium shaping effects, i.e. elongation and triangularity, on the lower hybrid
wave propagation and absorbtion in tokamaks, using the Solov'ev equilibrium. Numerical results show that, for midplane wave launching, increasing
elongation leads to down shift of the parallel wave number and central absorbtion, while for top and bottom launching, the effect tends to be the
opposite. Triangularity has the opposite effect of elongation, except for bottom launching, where the consequence of increasing triangularity is not
evident. The 2D mode structure of the lower hybrid wave is reconstructed by following the wave-packet propagation, including focusing/defocusing
effects. These studies have been extended to general geometries and equilibria given numerically, using a reference ITER equilibrium and the
corresponding ``fitted'' Solov'ev model. Good agreement was obtained for these cases, showing that it is reasonable to use a Solov'ev
parameterization for identifying interesting equilibrium parameter ranges to be investigated in more detail with full numerical equilibria.

As further application, we used the mixed WKB-full-wave method to explore the time-dependent 2D electrostatic ITG mode structure in the presence of a
(point-like) source term. While the parallel wave equation is calculated as an eigenvalue problem, the radial propagation is investigated using WKB
for solving the initial value radial envelope problem. The 2D global eigenmode is obtained by observing the time-asymptotic behavior of the
wave-packet, propagating between WKB turning points, while generic mode structures decay as the time increases because of phase mixing. For the sake
of simplicity, we assumed electrostatic fluctuations for both lower hybrid wave and ITG mode. However, the present wave-packet tracing method is
suitable for analyses of electromagnetic waves as well, including nonlinear wave interaction with zonal structures in the presence of a general
source and sink.  This perspective opens up the possibility of future applications of this present theoretical framework to a variety of interesting
topics.

%%%%%%%%%%%%%%%%%%%%%%%%%%%%%%%%%%%%%%%%%%%%%%%%%%%%%%%%%%%%%%%%%%%%

\section*{Acknowledgments}

This work is supported by the Chinese Scholarship Council (Grant No. 2010601100),  by Euratom
Communities under the contract of Association between EURATOM/ENEA, by the NSFC, Grant Nos.10975012, 40731056, National Basic Research Program of
China (973 Program), Grant No. 2008CB717803, and National ITER Program of China, Grant Nos. 2009GB105004 and 2009GB105006. Useful discussions with 
G. Calabr\`o, Z. Lin, Z. Qiu, G. Vlad, X. Wang and H. S. Zhang are also kindly acknowledged.
\appendix

\section{Derivation of the mode structure decomposition approach.}
\label{app:msd}

In this appendix and in appendix \ref{app:ballooning}, we discuss the detailed aspects of the mode structure decomposition approach, introduced in
section~\ref{sec:coordmsd}, while its connection with the well-known ``ballooning formalism"
(BF)~\cite{coppi77,lee77,glasser77,pegoraro78,connor78,connor79,hazeltine81,dewar81,dewar82} is analyzed  in appendix \ref{app:ballooning}.

As pointed out in section~\ref{sec:coordmsd}, the mode structure decomposition (MSD) approach was introduced in Refs.
[\!\!\citenum{cardinali03,zonca04}] for general analyses of wave/packet propagation in toroidal systems. Its validity, however, is more general and
can be extended to any three-dimensional systems with two periodic dependences, one of which is ignorable, i.e. corresponds to a symmetry of the
system itself. Fusion plasmas magnetically confined in helical systems and planetary magnetospheres are two obvious examples. With respect to Refs.
[\!\!\citenum{cardinali03,zonca04}], we provide here a more detailed and rigorous formulation of the MSD approach and the derivation of its
properties. Following notations introduced in section~\ref{sec:coordmsd}, we use a straight magnetic field line toroidal coordinate system
$(r,\theta,\zeta)$~\cite{roberts65,dewar83,cowley91,beer95,scott98}, where $r$ is a radial-like coordinate depending only on the magnetic flux
function $\psi$, $\theta$ and $\zeta$ are periodic angle like coordinates, chosen such that ${\bf B}\cdot {\bf \nabla} \zeta/{\bf B}\cdot {\bf
\nabla} \theta = q(r)$ and $\zeta$ corresponding to the symmetry of the system, the equilibrium magnetic field has the Clebsch representation ${\bf
B}= {\bf \nabla}\xi \times {\bf \nabla}\psi$ and $\xi = \zeta - q(r) \theta$~\cite{kruskal58}.

The formulation of the MSD approach and the derivation of its properties are based on the Poisson summation formula, which, in its most common form,
reads as
\begin{equation}
\sum_m e^{im\theta} = 2\pi \sum_m \delta (\theta - 2\pi m) \;\; . \label{eq:psf}
\end{equation}
Here and in the following, the summation is implicitly assumed to be on $m \in \mathbb{Z}$. In general, using the symmetry properties of the system,
the $\zeta$ dependence of a scalar function $f(r,\theta,\zeta)$, describing a generic fluctuating field, is decomposed as Fourier series
\begin{equation}
f(r,\theta,\zeta)= \sum_n e^{i n \zeta} f_n(r,\theta) \;\; , \label{eq:fourierdec}
\end{equation}
where the Fourier components $f_n(r,\theta)$ are independent in the linear limit. Note that time dependences are assumed implicitly for simplified
notations. The formulation of the MSD approach is based on the properties of sampling and periodization operators, which are related by the Fourier
transform due to the Poisson summation formula, which holds pointwise in the Schwartz space, but holds as well in a more general context under weaker
conditions~\cite{benedetto96}. The periodization operator $\hat f_n (r,\eta) \mapsto f_n(r,\theta)$ from ${\bf L}^1(\mathbb R)$ onto ${\bf
L}^1(\mathbb T)$ is expressed as
\begin{eqnarray}
f_n(r,\theta) & = & 2\pi \sum_\ell \hat f_n (r,\theta-2\pi\ell)\nonumber \\
& = & \sum_m e^{-im\theta} \int e^{im\eta} \hat f_n (r,\eta) d\eta \;\; , \label{eq:periodization}
\end{eqnarray}
where the second line is readily derived by use of equation (\ref{eq:psf}) and the $\eta$ integration is on $\mathbb R$. The periodization operator
sum converges absolutely a.e. on $\mathbb R$~\cite{benedetto96}, which is a useful property for the manipulations made hereafter. Note that the
possibility of introducing the periodization operator of equation (\ref{eq:periodization}) was adopted in Refs. [\!\!\citenum{connor79}] for
discussing the properties of the BF. Using equation (\ref{eq:periodization}), it is readily recognized that
\begin{eqnarray}
\left. \partial_r \right|_{\theta,\zeta} f_n(r,\theta) & \mapsto & \left. \partial_r\right|_{\eta,\zeta} \hat f_n(r,\eta)  \nonumber \\
\left. \partial_\theta\right|_{r,\zeta} f_n(r,\theta) & \mapsto & \left. \partial_\eta\right|_{r,\zeta} \hat f_n(r,\eta)  \nonumber \\
p(\theta) f_n(r,\theta) & \mapsto & p(\eta)  \hat f_n(r,\eta)  \;\; , \label{eq:mapping0}
\end{eqnarray}
where $p(\theta)$ is a generic periodic function. Once $\hat f_n(r,\eta)$ is known, $f_n(r,\theta)$ is uniquely determined by equation
(\ref{eq:periodization}). The opposite is not true. In fact, the Fourier transform of the periodization operator $\hat f_n (r,\eta) \mapsto
f_n(r,\theta)$ is the sampling operator $\varphi_n(r,x) \mapsto \left( \varphi_n(r,m) \right)_{m\in \mathbb Z}$, with
\begin{equation}
\varphi_n(r,x) = \int e^{i\eta x} \hat f_n (r,\eta) d \eta \;\; . \label{eq:fourierphi0}
\end{equation}
Equation (\ref{eq:periodization}) is readily rewritten as
\begin{eqnarray}
f_n(r,\theta) &=&  \sum_m e^{-im\theta}  \varphi_n(r,m)\\ \nonumber
 &=& \sum_m e^{-im\theta} \int e^{im\eta} \hat f_n (r,\eta) d\eta \;\; ; \label{eq:periodization1}
\end{eqnarray}
thus, the unique construction of
\begin{equation}
\hat f_n (r,\eta) = (2\pi)^{-1} \int e^{-i\eta x} \varphi_n(r,x) d x \;\;  \label{eq:fourierphi}
\end{equation}
is prevented from the fact we know the function $\varphi_n(r,x)$ via the sampling operator $\varphi_n(r,x) \mapsto \left( \varphi_n(r,m)
\right)_{m\in \mathbb Z}$. This fact was noted in Refs. [\!\!\citenum{cardinali03,zonca04}] and is not a problem, for the physical field
$f_n(r,\theta)$ can be always uniquely constructed from $\hat f_n(r,\eta)$, while it is not necessary and often not even useful to construct a unique
form of $\hat f_n(r,\eta)$, when the physical solution is known already. This viewpoint is identical to that adopted by Dewar and coworkers on the
construction of the inverse ballooning transformation~\cite{dewar83} (see appendix \ref{app:ballooning}). Possible constructions of $\varphi_n(r,x)$
and, therefore, of $\hat f_n (r,\eta)$ by equation (\ref{eq:fourierphi}), are obtained introducing the window function $w(x)$~\cite{hazeltine90},
which can be defined as a piecewise continuous function with maximum $w(0)=1$ at $x=0$ and vanishing everywhere outside the interval $(-1,1)$. In
fact, we have
\begin{equation}
\varphi_n(r,x)=\sum_m w(x-m) \varphi_n(r,m) \label{eq:interpolation}
\end{equation}
\begin{widetext}\begin{eqnarray}
\hat f_n (r,\eta) = \sum_m e^{-im\eta}  \varphi_n(r,m)  \frac{1}{2\pi} \int e^{-i\eta (x-m)} w(x-m) d x= g(\eta) f_n (r,\eta) \;\;
,\label{eq:fourierphi2}
\end{eqnarray}\end{widetext}
where $g(\eta)$ indicates the Fourier transform of the window function; i.e.
\begin{equation}
g(\eta) = \frac{1}{2\pi} \int e^{-i\eta x} w(x) d x \;\; . \label{eq:fourierg}
\end{equation}
Explicit examples of the functions $w(x)$ and $g(\eta)$ are given in Refs. [\!\!\citenum{cardinali03,zonca04}], with $g(\eta)$ satisfying the
condition
\begin{equation}
2\pi \sum_m g(\eta-2\pi m) = 1 \;\; , \label{eq:gnorm}
\end{equation}
which is straightforward consequence of its definition.

The periodization operator of equation (\ref{eq:periodization}) and its non-unique inverse of equation (\ref{eq:fourierphi2}) are of particular
interest when connected with the action of a generic bounded linear differential operator ${\cal L}(r,\theta;\partial_r,\partial_\theta) f_n
(r,\theta)$, which, considering equations (\ref{eq:mapping0}) and invoking the system periodicity in $\theta$, readily maps to ${\cal
L}(r,\eta;\partial_r,\partial_\eta) \hat f_n (r,\eta)$. In fact, using the absolute convergence a.e. in $\mathbb R$ of the periodization operator in
equation (\ref{eq:periodization}), we have by definition:
\begin{widetext}\begin{equation}
{\cal L}(r,\theta;\partial_r,\partial_\theta) f_n (r,\theta) = 2\pi \sum_\ell \int {\cal L}(r,\eta;\partial_r,\partial_\eta) \hat f_n (r,\eta) \delta
(\eta - \theta + 2\pi\ell) d \eta \;\; . \label{eq:diff0}
\end{equation}\end{widetext}
This equation shows that, if $\hat f_n (r,\eta)$ satisfies ${\cal L}(r,\eta; \partial_r,\partial_\eta) \hat f_n (r,\eta) = 0$, then $f_n (r,\theta)$
satisfies ${\cal L}(r,\theta; \partial_r,\partial_\theta) f_n (r,\theta) = 0$. Note that it was pointed out already in Refs.
[\!\!\citenum{connor79,dewar83}], discussing the properties of the BF, that  $f_n (r,\theta)$ obeys the same equation as $\hat f_n (r,\eta)$. Thus,
knowing the solution of a PDE in the $(r,\eta)$ space, allows us to construct uniquely the solution of the corresponding problem in the $(r,\theta)$
space. Section~\ref{sec:coordmsd} provides a discussion of the practical advantages coming from this property.

The opposite of this argument does not hold. i.e. given the solution of a PDE in the $(r,\theta)$ space, we cannot construct the solution of the
corresponding problem in the $(r,\eta)$ space, even admitting that the mapping $f_n (r,\theta) \mapsto \hat f_n (r,\eta)$ of equation
(\ref{eq:fourierphi2}) is not unique. With $\hat f_n (r,\eta) =  g(\eta) f_n (r,\eta)$, equation (\ref{eq:diff0}) becomes
\begin{widetext}\begin{eqnarray}
{\cal L}(r,\theta;\partial_r,\partial_\theta) f_n (r,\theta) & = & 2\pi \sum_\ell \int {\cal L}(r,\eta;\partial_r,\partial_\eta) \hat f_n (r,\eta)
\delta (\eta - \theta + 2\pi\ell) d \eta
\nonumber \\
& =  & 2\pi \sum_\ell {\cal L}(r,\eta;\partial_r,\partial_\theta) \left[ f_n (r,\theta) g(\theta - 2\pi\ell) \right] \;\; . \label{eq:diff1}
\end{eqnarray}\end{widetext}
Therefore, that ${\cal L}(r,\theta; \partial_r,\partial_\theta) f_n (r,\theta) = 0$ does not ensure that ${\cal L}(r,\eta; \partial_r,\partial_\eta)
\hat f_n (r,\eta) = 0$ but that $\sum_\ell {\cal L}(r,\eta;\partial_r,\partial_\theta) \left[ f_n (r,\theta) g(\theta - 2\pi\ell) \right]=0$, due to
the condition of equation (\ref{eq:gnorm}). Requesting that $\hat f_n (r,\eta)$ obeys the same equation as $f_n (r,\theta)$ yields the paradox that
$g(\eta)=1$~\cite{dewar83}, i.e. $w(x)=2\pi \delta(x)$, thus violating equation (\ref{eq:gnorm}). As mentioned above and noted in previous literature
on the BF~\cite{dewar83}, this is not a problem, for the only relevant fact is that the physical field $f_n(r,\theta)$ can be always uniquely
constructed from $\hat f_n(r,\eta)$ as solution of the corresponding PDE (see also appendix \ref{app:ballooning}).

In practical applications, it is often useful to move from straight magnetic field line toroidal coordinates $(r,\theta,\zeta)$ to Clebsch
coordinates  $(r,\theta,\xi)$~\cite{roberts65,dewar83,cowley91,beer95,scott98}, with $\xi = \zeta - q(r) \theta$~\cite{kruskal58}. It is readily
shown that equation (\ref{eq:fourierdec}) becomes
\begin{equation}
f(r,\theta,\xi)= \sum_n e^{i n \xi} F_n(r,\theta) \;\; , \label{eq:fourierdec2}
\end{equation}
so that, while periodicity in $\xi$ is maintained, periodicity in $\theta$ is substituted by $F_n(r,\theta+2\pi) = e^{2\pi i n q} F_n(r,\theta)$. In
$(r,\eta)$ space, the transform corresponding to using Clebsch coordinates is obtained by letting
\begin{equation}
\hat f_n (r,\eta) = e^{-inq\eta} \hat F_n(r,\eta) \;\; , \label{eq:hatbigf}
\end{equation}
corresponding to the periodization operator
\begin{eqnarray}
F_n(r,\theta) & = & 2\pi \sum_\ell e^{2\pi i \ell nq} \hat F_n (r,\theta-2\pi\ell)\nonumber \\
& = & \sum_m e^{i(nq-m)\theta} \int e^{i(m-nq)\eta} \hat F_n (r,\eta) d\eta \;\; . \label{eq:periodization2}
\end{eqnarray}
Considering the chain rules
\begin{eqnarray}
\left. \partial_r \right|_{\theta,\xi} & = & \left. \partial_r \right|_{\theta,\zeta} + q'(r) \theta \left. \partial_\zeta \right|_{r,\theta} \nonumber \\
\left. \partial_\theta \right|_{r,\xi} & = & \left. \partial_\theta \right|_{r,\zeta} + q(r) \left. \partial_\zeta \right|_{r,\theta} \nonumber \\
\left. \partial_\xi \right|_{r,\theta} & = & \left. \partial_\zeta \right|_{r,\theta} \;\; , \label{eq:chainrules}
\end{eqnarray}
and that, due to the definition in equation (\ref{eq:hatbigf}), the following mappings apply in the $(r,\eta)$ space
\begin{eqnarray}
\partial_r \hat f_n (r,\eta) & \mapsto & (\partial_r - in q'(r) \eta) \hat F_n (r,\eta) \nonumber \\
\partial_\eta \hat f_n (r,\eta) & \mapsto & (\partial_\eta - in q(r)) \hat F_n (r,\eta) \;\; , \label{eq:mapping1}
\end{eqnarray}
it is evident that ${\cal L}(r,\theta;\partial_r,\partial_\theta) f_n (r,\theta) \mapsto {\cal G}(r,\theta;\partial_r,\partial_\theta) F_n
(r,\theta)$ and that $F_n (r,\theta)$ obeys the same equation as $\hat F_n (r,\eta)$~\cite{connor79,dewar83} (cf. the above discussion for $f_n
(r,\theta)$ and $\hat f_n (r,\eta)$). In fact
\begin{widetext}\begin{eqnarray}
{\cal L}(r,\theta;\partial_r,\partial_\theta) f_n (r,\theta) & = & e^{-inq\theta} {\cal L}(r,\theta;\partial_r-inq'\theta,\partial_\theta-inq) F_n (r,\theta) \nonumber \\
& & \hspace*{-2cm} =  e^{-inq\theta} {\cal G}(r,\theta;\partial_r,\partial_\theta) F_n (r,\theta) \nonumber \\
& & \hspace*{-2cm} =  2\pi e^{-inq\theta} \sum_\ell \int e^{2\pi i \ell nq} {\cal L}(r,\eta;\partial_r-inq'\eta,\partial_\eta-inq) \hat F_n (r,\eta)
\delta (\eta - \theta + 2\pi\ell) d \eta
\nonumber \\
& & \hspace*{-2cm} = 2\pi e^{-inq\theta} \sum_\ell \int e^{2\pi i \ell nq} {\cal G}(r,\eta;\partial_r,\partial_\eta) \hat F_n (r,\eta) \delta (\eta -
\theta + 2\pi\ell) d \eta
 \;\; .
\label{eq:diff2}
\end{eqnarray}\end{widetext}
As concluded above for the functions $f_n (r,\theta)$ and $\hat f_n (r,\eta)$, solving the PDE $${\cal G}(r,\eta;\partial_r,\partial_\eta) \hat F_n
(r,\eta)=0$$ and constructing $F_n(r,\theta)$ from equation (\ref{eq:periodization2}) ensures that ${\cal G}(r,\theta;\partial_r,\partial_\theta) F_n
(r,\theta) = 0$. The opposite is not true; however, this fact does not pose issues of physical consistency~\cite{dewar83,cardinali03,zonca04}.

\section{From mode structure decomposition to ballooning transform}
\label{app:ballooning}

Equation (\ref{eq:periodization2}), which reflects the periodic structures underlying a generic fluctuation in a double periodic system that is
invariant under rotations in one of the two periodic coordinates, is the form used in the original works on the ballooning
representation~\cite{coppi77,lee77,glasser77,pegoraro78,connor78,connor79,hazeltine81,dewar81,dewar82} for analyzing its properties and elucidating
its usefulness in stability analyses of short wavelength (high-$n$) modes. In fact, when spatial scale separation between equilibrium profiles and
radial wavelength applies, such that $|\partial_r \hat F_n (r,\eta)| \ll |nq' \hat F_n (r,\eta)|$, then the radial structures of $e^{-inq\theta}
F_n(r,\theta)$ Fourier components essentially depend on $(nq-m)$, thus they are characterized by a near translational invariance. Actually, it has
been noted that the ballooning formalism (BF) is a convenient  formulation for treating the ``flute-like'' structures ($|nq-m|\ll 1$) that naturally
appear as fine radial scales (inertial/resistive layers) in resistive~\cite{pegoraro86} and ideal~\cite{newcomb90} MHD treatments of arbitrary mode
number fluctuations. For the same reason, the same approach has been adopted for analyzing in general the multiple spatial-scale nature of kinetic
MHD and Alfv\'enic modes excited by supra-thermal particle populations~\cite{chen88,biglari91,tsai93,chen94}. The connection of equation
(\ref{eq:periodization2}) with the BF and its possible generalizations has also been explored in situations where the local magnetic shear,
$rq'(r)/q(r)$, is vanishing, i.e. where the usual formulation of the BF does not apply~\cite{zonca02,connor04}.

The strength of the mode structure decomposition (MSD) approach, discussed in section~\ref{sec:coordmsd} and appendix~\ref{app:msd}, is that it
introduces in general the same formal properties of the BF, without any request of spatial scale separation or finite magnetic shear, and it reduces
readily to the usual BF in the appropriate limits where spatial scale separation applies~\cite{cardinali03,zonca04}. Applications of the MSD approach
are given in sections~\ref{sec:lhcold} and~\ref{sec:ITG}.

Equation (\ref{eq:periodization2}), with spatial scale separation, can be considered as definition of the ballooning transform. From previous
literature and appendix~\ref{app:msd}, we know that solving the PDE $${\cal G}(r,\eta;\partial_r,\partial_\eta) \hat F_n (r,\eta)=0$$ and
constructing $F_n(r,\theta)$ from equation (\ref{eq:periodization2}) ensures that ${\cal G}(r,\theta;\partial_r,\partial_\theta) F_n (r,\theta) = 0$.
Appendix~\ref{app:msd} also discussed why, given $F_n(r,\theta)$, is generally not possible to uniquely construct $\hat F_n (r,\eta)$. Although this
fact does not pose issues of physical consistency, the question of the uniqueness of the inverse ballooning transform has attracted significant
interest~\cite{hazeltine81,dewar83,hazeltine90}. The crucial ingredient for the construction of the inverse ballooning
transform~\cite{hazeltine81,hazeltine90} is the (infinite) separation of scales, which ``identifies'' radial coordinate and parallel wave-vector,
i.e. introduces the notion that radial and parallel coordinates are Fourier transform duals or conjugate variables. In general, when spatial scale
separation does not apply, the function $\varphi_n(r,x)$, entering the inverse transform definition of equation (\ref{eq:fourierphi}), is known only
via the sampling operator $\varphi_n(r,x) \mapsto \left( \varphi_n(r,m) \right)_{m\in \mathbb Z}$.

Following Hazeltine and Newcomb~\cite{hazeltine90}, the ballooning transform and its inverse are given by [cf. equation(\ref{eq:periodization2})]
\begin{eqnarray}
f_n(r,\theta) & = & f_n (x,\theta;r) \\ \nonumber
& = & 2\pi \sum_\ell e^{ i  nq (2\pi \ell - \theta)} \hat F_n (r,\theta-2\pi\ell) \;\; , \label{eq:ballooning} 
\end{eqnarray}
\begin{eqnarray}
\hat F_n (r,\theta) & = & \frac{1}{2\pi} \int g(s) e^{i(x+s)\theta} f_n (x+s,\theta;r) ds \;\; . \label{eq:inverse}
\end{eqnarray}
Note that, here, spatial scale separation has been indicated explicitly by the notation $f_n(r,\theta) = f_n (x,\theta;r)$, with $x = nq$.
Furthermore, the function $g$ is the Fourier transform, given in equation~(\ref{eq:fourierg}), of the window function $w$ defined in
appendix~\ref{app:msd}. In the work by  Hazeltine and Newcomb~\cite{hazeltine90}, the function $g(s)=(\pi s)^{-1} \sin (\pi s)$. In the following, we
show that the inverse transform, given in equation~(\ref{eq:inverse}), applies for any generic form of $g(s)$ given in equation~(\ref{eq:fourierg}).

Since (infinite) spatial scale separation applies, we have that $e^{i\ell \theta} f_n (nq+\ell,\theta;r) =  f_n (nq,\theta;r)$. In fact,
\begin{eqnarray}
f_n (nq+\ell,\theta;r) & = & \sum_m e^{-im\theta} f_{nm} (nq+\ell - m;r) \nonumber \\
& = & e^{-i\ell \theta} \sum_k e^{-i k \theta} f_{nm} (nq - k;r) \\ \nonumber
& = & e^{-i\ell \theta} f_n (nq,\theta;r) \;\; . \label{eq:translation}
\end{eqnarray}
As in Ref. [\!\!\citenum{hazeltine90}], we substitute equation~(\ref{eq:inverse}) into equation~(\ref{eq:ballooning}) and show that we have an
identity:
\begin{widetext}\begin{eqnarray}
f_n (x,\theta;r) & = & \sum_\ell e^{ i  nq (2\pi \ell - \theta)} \int g(s) e^{i(x+s) (\theta - 2 \pi \ell) } f_n (x+s,\theta;r) ds \nonumber \\
& = &  \int g(s) e^{is \theta} \sum_m \delta (s-m) f_n (x+s,\theta;r) ds \nonumber \\
& = &  \int g(s) \sum_m \delta (s-m) f_n (x,\theta;r) ds \nonumber \\
& = & \left( \sum_\ell w(2\pi\ell) \right)  f_n (x,\theta;r) =  f_n (x,\theta;r) \;\; . \label{eq:consistency1}
\end{eqnarray}\end{widetext}
Here, we have used equation~(\ref{eq:translation}) in the second line and, in the fourth line, the fact that $\sum_\ell w(2\pi\ell) = w(0) = 1$ by
definition of the window function $w$, given in appendix~\ref{app:msd}. Note that, with the special choice of $g(s)=(\pi s)^{-1} \sin (\pi s)$, made
in Ref. [\!\!\citenum{hazeltine90}], the identity is proved in the second line, since $\sum_m g(s) \delta (s-m) = \delta(s)$. An identity is also
found when substituting equation~(\ref{eq:ballooning}) into equation~(\ref{eq:inverse}).
\begin{eqnarray}
&\hat F_n (r,\theta)  =  \sum_\ell \int g(s) e^{i2\pi\ell(x+s)} \hat F_n (r,\theta-2\pi\ell) ds& \nonumber \\
& =  \sum_\ell w(-2\pi\ell) e^{i2\pi\ell x} \hat F_n (r,\theta-2\pi\ell) = \hat F_n (r,\theta) \;\; . &\label{eq:consistency2}
\end{eqnarray}
Thus, the results reported here generalize those given by Hazeltine and Newcomb~\cite{hazeltine90} and show that the most general pair of ballooning
transformation and its inverse are provided by equations~(\ref{eq:ballooning}) and~(\ref{eq:inverse}).

The general discussion of the connection of the MSD approach with the standard BF is completed by the proof that, if $f_n
(r,\theta)=e^{-inq\theta}F_n(r,\theta)$ satisfies ${\cal L}(r,\theta; \partial_r,\partial_\theta) f_n (r,\theta) = 0$, then $\hat f_n (r,\eta)$
satisfies ${\cal L}(r,\eta; \partial_r,\partial_\eta) \hat f_n (r,\eta) = 0$. The opposite is always true even in the more general MSD approach, as
shown by equation~(\ref{eq:diff0}). Given the inverse transform of equation~(\ref{eq:inverse}), the following mappings are readily shown:
\begin{eqnarray}
\partial_r \hat f_n (r,\theta) & \mapsto & \partial_r  f_n (x+s,\theta;r) \nonumber \\
(\partial_\theta + i x ) \hat f_n (r,\theta) & \mapsto & \left( i (s+x) + \partial_\theta\right) f_n (x+s,\theta;r) \;\; , \label{eq:mapping2}
\end{eqnarray}
where the second line represents the mapping of the parallel derivative. The mapping of the poloidal derivative is trivially obtained from the
spatial scale separation argument, for which $\partial_\theta f_n(x,\theta;r) \simeq - ix f_n(x,\theta;r)$. Thus
\begin{eqnarray}
\partial_\theta  \hat f_n (r,\theta)  & \mapsto &  \left( i s + \partial_\theta\right) f_n (x+s,\theta;r) \\ \nonumber
& \simeq & - i x f_n (x+s,\theta;r) \;\; . \label{eq:mapping3}
\end{eqnarray}
Equations (\ref{eq:mapping2}) and~(\ref{eq:mapping3}) are the formal proof that, if $f_n (nq,\theta;r)$ and $\hat F_n (r,\theta)=e^{inq\theta}\hat
f_n(r,\theta)$ are related by the transforms of equations~(\ref{eq:ballooning}) and~(\ref{eq:inverse}), ${\cal L}(r,\theta;
\partial_r,\partial_\theta) f_n (r,\theta) = 0$ implies  ${\cal L}(r,\eta; \partial_r,\partial_\eta) \hat f_n (r,\eta) = 0$.

\iffalse
\section{Solov'ev equilibrium}
\label{app:solovev} \fi
%%%%%%%%%%%%%%%%%%%%%%%%%%%%%%%%%%%%%%%%%%%%%%%%%%%%%%%%%%%%%%%%%%%%

%\section*{BibTeX bibliography usage}

%This section should be removed. It is used in this example just to demonstrate the use of BibTeX, e.g. citing as usual \cite{connor78}, References
%[\!\!\citenum{glasser77,lee77}], and as usual again\cite{cardinali03}.

% The following makes use of BibTeX
%
%\bibliographystyle{unsrt}
\bibliography{zlubib}

%%%%%%%%%%%%%%%%%%%%%%%%%%%%%%%%%%%%%%%%%%%%%%%%%%%%%%%%%%%%%%%%%%%%

\newpage

\end{document}